\documentclass[reprint,aps,prx,nofootinbib]{revtex4-2}
\usepackage[utf8]{inputenc}
\usepackage{mathrsfs}
\usepackage{natbib}
\usepackage{amsmath}
\usepackage{amsfonts}
\usepackage{amssymb}
\usepackage{graphicx}
\usepackage{hyperref}
\usepackage[usenames,dvipsnames]{color}
\usepackage{mathrsfs}

\newcommand{\abs}[1]{\left\vert#1\right\vert}
\newcommand{\set}[1]{\left\{#1\right\}}

\newcommand{\eps}{\varepsilon}

\newcommand{\del}{\boldsymbol{\nabla}}
\newcommand{\arccosh}{\mathrm{arccosh}}
\newcommand{\zhat}{\mathbf{\hat{z}}}
\newcommand{\yhat}{\mathbf{\hat{y}}}
\newcommand{\xhat}{\mathbf{\hat{x}}}

\newcommand{\phihat}{\boldsymbol{\hat{\phi}}}
\newcommand{\aquitania}{\mathcal{A}}
\newcommand{\belgica}{\mathcal{B}}
\newcommand{\celtica}{\mathcal{C}}
\newcommand{\pois}[1]{\set{#1}}
\newcommand{\ham}{\mathscr{H}}
\newcommand{\bin}{\textrm{bin}}

\newcommand{\tide}{\textrm{tide}}
\newcommand{\scri}{\mathscr{I}}

\def\aj{\textrm{AJ}}                   
\def\actaa{\ref@jnl{Acta Astron.}}      
\def\apj{\textrm{ApJ}}                 
\def\apjs{\textrm{ApJS}}               
\def\apss{\textrm{Ap\&SS}}             

\def\mnras{\textrm{MNRAS}}             
\def\sovast{\textrm{Soviet~Ast.}}      
\def\nat{\textrm{Nature}}              

\begin{document}

\title{An Analytical, Statistical Approximate Solution for Dissipative and non-Dissipative \\
Binary-Single Stellar Encounters}

\author{Yonadav Barry Ginat and Hagai B. Perets}
\affiliation{Faculty of Physics, Technion -- Israel Institute of Technology, Haifa, 3200003, Israel}%
\email{ginat@campus.technion.ac.il}%
\email{hperets@physics.technion.ac.il}

\begin{abstract}
We present a statistical approximate solution of the bound, non-hierarchical three-body problem, and extend it to a general analysis of encounters between hard binary systems and single stars. Any such encounter terminates when one of the three stars is ejected to infinity, leaving behind a remnant binary; the problem of binary-single star-scattering consists of finding the probability distribution of the orbital parameters of the remnant binary, as a function of the total energy and the total angular momentum. Here, we model the encounter as a series of close, non-hierarchical, triple approaches, interspersed with hierarchical phases, in which the system consists of an inner binary and a star that orbits it -- this turns the evolution of the entire encounter to a random walk between consecutive hierarchical phases. We use the solution of the bound, non-hierarchical three-body problem to find the walker's transition probabilities, which we generalise to situations in which tidal interactions are important. Besides tides, any dissipative process may be incorporated into the random walk model, as it is completely general. Our approximate solution can reproduce the results of the extensive body of past numerical simulations, and can account for different environments and different dissipative effects. Therefore, this model can effectively replace the need for direct few-body integrations for the study of binary-single encounters in any environment. Furthermore, it allows for a simply inclusion of dissipative forces typically not accounted for in full N-body integration schemes.
\end{abstract}

\maketitle

\section{Introduction}
The three-body problem -- that of the interaction of three gravitating objects -- is one of the oldest physical and astrophysical problems, and has been studied for centuries; solving it requires an understanding of the outcomes of encounters between binary systems and single stars. Such encounters play a key r\^{o}le in the evolution of binary systems, both in dense environments such as globular or nuclear clusters \citep{Binney,HeggieHut2003}, and in the field, too \citep{Per+12,Mic+20}. These encounters can result in the formation of compact binaries, and consequently engender a plethora of physical phenomena, ranging from mergers of stars and compact objects, to the production of electromagnetic and gravitational-wave transients, such as type Ia supernov\ae, short gamma-ray bursts, \emph{et cetera}. Currently, full numerical integrations of few-body systems are required in order to characterise the results of such interactions. However, the large number of such few-body simulations needed for one cluster comes with a significant computational cost; any change in the basic aspects of the problem (e.g. different masses, binary configuration etc.) necessitates a new set of simulations; and likewise different environments (e.g. external, outer potentials in clusters) require different sets of simulations (and in many cases are not even consistently accounted for). Moreover, realistic physical processes that may affect the outcomes, such as dissipative forces (e.g. tidal interactions, or gravitational-wave dissipation) acting on the interacting stars are rarely incorporated into simulations of the evolution of binary systems through binary-single encounters. Finally, such simulations, by their very nature, provide neither any explanation for their outcomes, nor any direct insight into them, and are used, to some extent, as black-box ingredients in simulations of large-scale systems. Despite much progress in the analytical understanding of the problem, the full solution for the outcomes of non-hierarchical three-body systems, and in particular the general understanding of binary-single encounters and their long-term end-states under realistic conditions remains unsolved. Furthermore, accounting analytically for dissipative aspects of this problem was not done previously, to the best of our knowledge.

Here we provide a full analytical, statistical model that solves the non-hierarchical three-body problem in the limit of encounters with hard binaries. We show how our approach can account for the long-term evolution of a non-hierarchical triple system and characterise it as it goes through consecutive close binary-single encounters until the system is destroyed. Our solution provides the detailed cross-section for the final outcomes, including both cases of ejection of one of the components or the collision/merger of two of the components, as well as, most importantly, the characterisation of the final remnant binary properties. Our approach can also account for different spatial cut-offs due to external perturbations. Last, but not least, it can generally account for dissipative forces -- we exemplify its application for dissipative tidal forces during binary-single encounters. Thus, the use of our model, which is based on a random-walk approach, can potentially replace the need for direct integration of binary-single encounters and provide a general, robust tool for the understanding the outcomes of binary-single encounters.

Binary-single encounters can generally be divided between the cases of encounters with hard or soft binaries, i.e. according to the ratio between the binary binding energy and the kinetic energy of the incoming third star. When the binary star is wide -- when its binding energy is considerably smaller than the typical kinetic energy of a star in the host cluster, $k_BT$ (where $T$ is proportional to the squared velocity dispersion of the cluster), the encounter is well-described by a composition of two two-body interactions \citep{Heggie1975,Binney,ValtonenKarttunen2006}. In the case where the binary is `hard', i.e. when its binding energy is much larger than $k_BT$, an analytical description of the encounter is more difficult.
If the encounter stays hierarchical, it may be treated by means of perturbation theory \citep{Heggie1975}, but if it doesn't, it becomes a so-called `resonant encounter', where there are phases when all three bodies are close to each other, and the evolution is inherently chaotic. One could still find some analytical insight by investigating the phase-space evolution of a set of various initial conditions during such a close encounter. For example, it has been shown at the early 20-th century that if the system starts as a binary and an unbound third star, then, eventually, it will end up with at least one unbound star (apart from a set of measure zero); see \citet[\S\S 2.4, 2.6]{Arnoldetal2006} for a review of this theorem and related ones.

Here we wish to present an analytical, statistical model for the evolution of such encounters with hard binaries. We will endeavour to derive a probability distribution function for the possible end-states. The initial binary is hard, so the total energy of the system, $E$, must be negative, whence an end-state with three stars unbound is forbidden -- the set of all possible end-states is therefore the set of all possible final binary configurations, multiplied by the set of configurations of the ejected star (which, of course, may be a different star from the initial unbound one -- see \S \ref{sec:masses} below). We derive a distribution function, $f_\bin(E_\bin,\mathbf{S}|E,\mathbf{J})$, for the resultant binary to have energy $E_\bin$ and angular momentum $\mathbf{S}$, given that the total energy is $E$ and that the total angular momentum is $\mathbf{J}$.

Let $m_1$, $m_2$, $m_3$ denote the masses of the three stars, which we take to be of similar magnitude, and let $M = m_1 + m_2 + m_3$ be the total mass. Further let the subscript `$\bin$' denote any quantity related to the binary, such as its total mass $m_\bin$, the reduced mass of its two components $\mu_\bin$, \emph{et cetera}. Likewise, let a subscript $s$ denote quantities pertaining to the ejected star, like $\mu_s = m_sm_\bin/M$. We denote the total energy of the triple by $E$ or sometimes by $E_0$, and its total angular momentum by $\mathbf{J}$. More explicitly then, the problem we investigate is the following one: we consider a single star with velocity $v_0$ at infinity, which is scattered on a hard binary, such that the total energy is $E$ and the total angular momentum is $\mathbf{J}$. As the binary is hard, $E \approx - Gm_am_b/(2a_0)$, where $a_0$ is the initial semi-major axis of the binary, consisting, initially, of stars $a$ and $b$. For a given impact parameter $b$ of the single star, relative to the binary centre-of-mass (as well as all the initial anomalies of the binary), it is possible, in principle, to predict the outcome of the encounter, i.e., the final values of $E_\bin$, $\mathbf{S}$ and $E_s$. However, due to the chaotic nature of the close three-body interaction, doing so is impracticable. One could study the problem either by performing numerical simulations (see, e.g., refs. \cite{Saslawetal1974,Hills1975,HillsFullerton1980,Anosova1986,AnosovaOrlov1986,Hills1989,Hills1992,HeggieHut1993,Hut1993,SigurdssonPhinney1993,Mikkola1994,Samsingetal2014,
LeighWegsman2018,Manwadkaretal2020}), or, statistically, using various analytical approximations. Such analytical treatments presuppose that the results of such numerical scattering experiments may be treated as random variables, drawn from some distribution; in effect, it reduces the problem to finding that distribution. Numerical simulations have shown that the encounter proceeds as a sequence of many close triple approaches, after each of which a single star is ejected \citep{Anosova1986,AnosovaOrlov1986,Samsingetal2014}: if the star is still bound to the other binary, it returns eventually, whereupon a new close triple approach ensues, and so on, until the single star is ejected to infinity with positive energy.

There are two dominant analytical approaches in the literature (to our knowledge): one, initiated by \citeauthor{Monaghan1976a} in \cite{Monaghan1976a}, relies on chaotic mixing during the close triple approach to argue that the distribution is ergodic in the relevant part of the system's phase space. This approach was refined later by various authors, e.g., refs. \cite{Monaghan1976b,NashMonaghan1978,ValtonenKarttunen2006,Kol2020}, to account more accurately for angular momentum conservation, until \citet{StoneLeigh2019} succeeded in doing so fully, for the unbound case.

The other approach, which dates back to \citet{Heggie1975}, was to draw upon the principle of detailed balance to deduce what the outcome distribution, $f_\bin(E_\bin,\mathbf{S}|E,\mathbf{J})$, must be in order for the number of bound triples formed by such encounters to be fixed, in a cluster in thermal equilibrium (see, e.g., refs. \cite{Heggie1975,HeggieHut1993,HeggieHut2003}). Various authors also attempted to account, in simulations, not only for the Newtonian interaction of three point-particles, but rather include other physical effects, such as collisions \citep{HutInagaki1985,SigurdssonPhinney1993}, and more recently also gravitational-wave emission and tidal dissipation \citep{Samsingetal2014,Samsingetal2017}. The reader is referred to chapters 7-8 of ref. \cite{ValtonenKarttunen2006} for a review of some of the work on binary-single scattering.

The contribution of this work to the understanding of binary-single encounters is threefold: first, we present the first (to our knowledge) closed-form statistical approximate solution of the bound, non-hierarchical three-body problem that takes both energy and angular momentum conservation into account, which complements the solution of the unbound case of ref. \cite{StoneLeigh2019}; using the results presented here, one can compute the outcome distribution of each intermediate close approach, not only of the final one.\footnote{While we were working on this paper, ref. \cite{Kol2020} provided an alternative approximation to the solution of the non-hierarchical three-body problem, which can be continued analytically to the bound case. This solution contains an unknown `emissivity' multiplicative factor, which has not yet been computed, and therefore does not constitute a closed-form solution.} Secondly, we show how to derive the solution both using detailed balance arguments and using ergodic arguments, thereby unifying the two approaches. That they give the same result is hardly surprising, since the principle of detailed balance is essentially a statement about phase-space volumes. Thirdly, and most significantly, we elevate our solution of the bound case to a random-walk model of the entire encounter, in which the binary actions and the three-body conserved quantities perform a random walk, whose transition probabilities are related to $f_\bin(E_\bin,\mathbf{S}|E,\mathbf{J})$.
This random walk model is extremely general, and can incorporate many physical processes beyond the Newtonian gravitational three-body interaction and in addition to it. Here we provide as an example the important case of dissipation due to tidal forces, but our approach can be generalized to any other type of additional perturbations and physical processes, which will be done in future work.

The structure of this paper is as follows: we start with an explicit calculation of $f_\bin(E_\bin,\mathbf{S}|E,\mathbf{J})$ using ergodic arguments, in \S \ref{sec: phase space integration}, while \S \ref{sec: detailed balance} presents a computation of the same function using the principle of detailed balance. In \S \ref{sec:masses} and \S \ref{sec:marginal energy distribution} we discuss two marginal distributions: the probability function of the ejected star's mass, and the marginal energy distribution. Our paper culminates in \S \ref{sec:random walks}, where we move on to present the random walk model in its full generality. In \S \ref{sec:comparison with simulations -- marginal distributions} we compare the results of \S\S \ref{sec:masses} and \ref{sec:marginal energy distribution} to past numerical simulations, and in \S \ref{sec:numerical simulations} we apply the random walk model to tidal dissipation, while also comparing it with relevant simulations. All of our results mesh well with the simulations.

\section{Cross-Section Via Phase-Space Integration}
\label{sec: phase space integration}

All the system's phase-space is divided into three parts: one where the system is hierarchical, but the outer body is unbound; one where the system is still hierarchical but the outer body is bound; and a chaotic region in which all three bodies interact closely. Let us denote them by $\mathcal{A},\mathcal{B}$ and $\mathcal{C}$, respectively. The latter is taken to be the collection of points where the maximum relative separation between any one body and the centre-of-mass of the other two is no more than some value $R$, which is defined below. Then, if chaos is sufficiently strong to lead to phase-space mixing inside $\celtica$, the cross-section of the outcome of a close approach is simply an integral over $\celtica$, as assumed in previous studies \citep{Monaghan1976a,Monaghan1976b,NashMonaghan1978,ValtonenKarttunen2006,StoneLeigh2019}:\footnote{Strictly speaking, one needs to normalise this integral so that it has the correct dimensions. This can be done by working in the appropriate system of units, and by multiplying the integral by functions that are symmetric in all three stars, and is therefore unimportant for the analysis presented in this paper.}
\begin{equation}
\begin{aligned}
  \sigma & = \left(\prod_{i=1}^{3}\int_\celtica \mathrm{d}^3\mathbf{r}_i\mathrm{d}^3\mathbf{p}_i\right)\delta(E-\ham)\delta(\mathbf{J}- \sum_{j=1}^{3}\mathbf{r}_j\times\mathbf{p}_j)\\ &
  \times \delta(\mathbf{P}_\textrm{CM} - \sum_{j=1}^{3}\mathbf{p}_j),
\end{aligned}
\end{equation}
where $\mathbf{r}_j$ and $\mathbf{p}_j$ are the position vector and the momentum of the $j$-th particle, $\ham$ is the Hamiltonian, and $E$, $\mathbf{P}_\textrm{CM}$ and $\mathbf{J}$ are the total energy, momentum and angular momentum, respectively.
This equation is equivalent to the statement that the probability of the system exiting $\celtica$ through a point $w$ is actually independent of $w$; we give a heuristic argument for this statement in appendix \ref{sec: chaos diffusion}.

One can calculate this integral by transforming to the co-ordinate system of a binary and a lone star orbiting its centre-of-mass, which amounts to demanding that the system becomes hierarchical when it leaves $\celtica$; this requirement is what defines $R$ below. Working in the centre-of-mass frame:
\begin{equation}\label{eqn: sigma}
  \sigma = \int \mathrm{d}^3\mathbf{r}_s\mathrm{d}^3\mathbf{p}_s\mathrm{d}^3\mathbf{r}_\bin\mathrm{d}^3\mathbf{p}_\bin\delta(E-E_\bin - E_s)\delta(\mathbf{J}- \mathbf{L} - \mathbf{S}),
\end{equation}
where $E_\bin = -Gm_1m_2/2a_\bin$ is the binary energy, and
\begin{equation}
  E_s = \frac{\abs{\mathbf{p}_s}^2}{2\mu_s} - \frac{G(m_1+m_2)m_3}{\abs{\mathbf{r_s} - \mathbf{r}_{\textrm{cm},\bin}}} = \pm\frac{G(m_1 + m_2)m_3}{2a_s},
\end{equation}
where the sign is negative/positive when the system goes into $\aquitania$/$\belgica$ respectively. The spin of the binary is $\mathbf{S}$ and the angular momentum of the third body about the binary is $\mathbf{L}$. Below the integral over the remaining binary phase-space is shortened to $\int_\bin$.

As there are three stars, with possibly different masses, there are three distinct outcomes, depending on which of the three bodies ends up being ejected. As the original binary was hard, the final binary must also be hard. Formally, we weigh the binding energies between each of the three pairs, and the one whose binding energy is considerably lower (i.e. more negative) is the remaining binary. This implies that the total cross-section splits into a sum of three cross-sections
\begin{equation}
  \sigma = \sigma_1 + \sigma_2 + \sigma_3,
\end{equation}
where $\sigma_i$ is the cross-section for a break-up with star $i$ ejected as the lone star in the hierarchical system. This also gives an infra-red cut-off for $E_\bin$ naturally, forcing it to be less than $E/3$, for if it were more, one of the other pairs would actually be harder.
Below we calculate such a $\sigma_i$; because all three are symmetric, we ignore this complication for now, and return to it in \S \ref{sec:masses}.

\subsection{$\celtica$ in Angle-Action Variables}
\label{subsec: Celtica in angle-action variables}
Performing this integral in angle-action variables is the best way to proceed \citep{StoneLeigh2019}, but first it is essential to determine the integration region explicitly: in the initial encounter, the third body must come close enough to the binary, such that its pericentre distance is of the same order as $R$. This implies that
\begin{equation}\label{eqn: close encounter condition}
  \begin{aligned}
  & a_s(1-e_s) \lesssim R ~\textrm{for an elliptic orbit, or} \\ & a_s(e_s - 1)\lesssim R ~\textrm{for a hyperbolic orbit}.
  \end{aligned}
\end{equation}
Equation \eqref{eqn: close encounter condition} needs to be supplemented by another condition, which says that while the system becomes hierarchical, the lone star's apoapsis is sufficiently larger than $R$. This condition is automatically verified for a hyperbolic orbit (where the lone star escapes and the preceding close encounter is the final one), but for an elliptic orbit (where the lone star eventually returns), one must have
\begin{equation}\label{eqn: hierarchy condition}
  a_s(1+e_s) \gtrsim \eta R,
\end{equation}
for $\eta\approx 5$ (other values are fine, too).

Let us discuss how $R$ is related to the binary parameters: $R$ is defined as the critical distance between the would-be ejected star to the centre-of-mass of the other two stars, where the problem becomes hierarchical. Conversely, for a hierarchical triple, one might write the Hamiltonian (in the centre-of-mass frame) as
\begin{equation}
  \ham = E_\bin + E_s - \frac{G}{r_s}\sum_{n=2}^{\infty}M_n\frac{r_\bin^n}{r_s^n}P_n(\cos\Phi),
\end{equation}
where
\begin{equation}
  M_n = \frac{m_1m_2m_3\left(m_a^{n-1} - (-m_b)^{n-1}\right)}{m_\bin^n},
\end{equation}
$m_a$ and $m_a$ are the inner binary's masses, $r_\bin$ is the distance between them, $\Phi$ is the angle between $\mathbf{r}_\bin$ and $\mathbf{r}_s$, and $P_n$ is the $n$-th Legendre polynomial. We refer the reader to refs. \cite{ValtonenKarttunen2006,Naoz2016} for details and for more references on the hierarchical three-body problem. Now, the triple ceases to be hierarchical if the multipole series becomes as large as the leading order term in the Hamiltonian, namely, as big as the energy $E = E_\bin + E_s$. The leading term in this series is the quadrupole (for non-extreme mass ratios). Approximating $r_\bin^2P_2(\cos\Phi) \approx a_\bin^2$, we find
\begin{equation}
  R \leq \beta \left(\frac{G\mu_\bin\mu_sM}{m_\bin\abs{E}}\right)^{1/3}a_\bin^{2/3},
\end{equation}
where $\beta$ is a constant of order unity. If the masses are un-equal, this formula might lead to an over-estimate, since then it is possible that an exchange of a light star with a heavy star would yield $R \gg a_\bin$, in which case the problem is still quite visibly hierarchical. To account for that, we define $R$ as
\begin{equation}\label{eqn: R definition}
  R = \beta \min\set{\left(\frac{G\mu_\bin\mu_sM}{m_\bin\abs{E}}\right)^{1/3}a_\bin^{2/3},a_\bin},
\end{equation}
where, again, $\beta \geq 1$ is of order unity; the cross-section $\sigma$ depends on $\beta$ only weakly (cf. ref. \cite{StoneLeigh2019} for the unbound case). In future work we will explore the consequences of the existence of this threshold for the stability of hierarchical triples. If the reader is concerned about the crude approximation of the Legendre polynomial, we offer a more refined one, and test both in appendix \ref{appendix:inclination}, but we recommend that it be read only after the next section.

Both conditions \eqref{eqn: close encounter condition} and \eqref{eqn: hierarchy condition} may be translated into conditions on the angular momentum $L$ (they are, evidently, independent of its direction). Using energy conservation one may express $E_s$ in terms of $E_\bin$, whence conditions \eqref{eqn: close encounter condition} and \eqref{eqn: hierarchy condition} may be written as $L \leq A(E_\bin)$. Let us start with the apoapsis condition:
First, if $\eta R < a_s$, then this condition is fulfilled automatically. If not, it simplifies to
\begin{equation}\label{eqn: apoapsis}
  L^2 \leq \mu_s^2GM \eta R\left(2 - \frac{\eta R}{a_s}\right).
\end{equation}
As the left-hand-side is non-negative, this implies that $\eta R \leq 2 a_s$. Therefore $a_s$ can either be more than $\eta R$, or more than $\eta R/2$, so in both cases more than $\eta R/2$. Thus, one has a restriction on the energy-difference $\abs{E - E_\bin}$, \emph{viz.}
\begin{equation}\label{eqn: energy restriction bound}
  \abs{E - E_\bin} \leq \frac{Gm_\bin\mu_s}{\eta R}.
\end{equation}

The elliptic periapsis condition is satisfied trivially if $a_s < R$, but if $\eta >2$ this cannot be the case. Otherwise, one has
\begin{equation}\label{eqn: periapsis bound}
  L^2 \leq \mu_s^2GMR\left(2 - \frac{R}{a_s}\right).
\end{equation}
Which of conditions \eqref{eqn: apoapsis} and \eqref{eqn: periapsis bound} is more stringent depends on the masses, and on $a_s$.  For the unbound, hyperbolic case, the periapsis condition is equivalent to
\begin{equation}\label{eqn: periapsis unbound}
  L^2 \leq \mu_s^2GMR\left(2 + \frac{R}{a_s}\right);
\end{equation}
note that as a consequence of the plus sign inside the brackets here, there is no restriction on $a_s$ in this case, and it may be as small as one pleases.

The function $A(E_\bin)$ is therefore defined as
\begin{widetext}
\begin{equation}
  A(E_\bin)^2 = \mu_s^2GMR\begin{cases}
                  \min\set{2 - \frac{R}{a_s(E_\bin)},\eta\left(2 - \frac{\eta R}{a_s(E_\bin)}\right)}, & \mbox{if } E < E_\bin \\
                  2 + \frac{R}{a_s(E_\bin)}, & \mbox{otherwise}.
                \end{cases}
\end{equation}
\end{widetext}

The approximation made here, which includes a separation of phase space into the three regions $\aquitania$, $\belgica$ and $\celtica$, and the different treatment of each, inadvertently induces some uncertainty. We attempt to gauge it in appendix \ref{appendix:error estimates}, but we urge the reader to peruse \S \ref{sec: detailed balance}, \S\ref{sec:masses} and \S \ref{sec:marginal energy distribution} before turning to this appendix.

\subsection{Cross-Section Calculation}
Now one may turn to performing the integration in equation \eqref{eqn: sigma}. The goal is to find the final distribution of binary spin and energy, of the remaining binary, after the system stops being chaotic. Thus, one has to integrate over all of the lone star's phase-space, as well as the angle variables of the binary. First, though, note that there are two possibilities for each encounter: either $E_s < 0$, or $E_s > 0$ after the encounter ($E_s = 0$ has zero measure). So $\sigma = \sigma_\textrm{bd} + \sigma_\textrm{ubd}$, where each cross-section pertains to each possible sign of $E_s$.\footnote{The triple is in a cluster, so it is possible that the star escapes even with $E_s < 0$, but that does not affect the calculation at present.}
$\sigma_\textrm{ubd}$ has been calculated by \citet{StoneLeigh2019}. They obtained
\begin{widetext}
\begin{equation}\label{eqn: Stone and Leigh 2019}
  \sigma_\textrm{ubd} \propto \int  \frac{\mathrm{d}E_\bin \mathrm{d}J_a\mathrm{d}J_b}{\abs{E - E_\bin}^{3/2}\abs{E_\bin}^{3/2}\abs{\mathbf{J}-\mathbf{S}}}\left[ \sqrt{\frac{R^2}{a_s^2} + \frac{2R}{a_s} + 1 - e_s^2} - \arccosh\left(\frac{R + a_s}{a_se_s}\right)\right].
\end{equation}

The other cross-section, $\sigma_\textrm{bd}$, may be calculated using Delaunay variables for both the emergent binary and the binary formed by the star and the inner binary. We do so below; this is meaningful given that the whole encounter proceeds as a series of consecutive close approaches, each one having a cross-section $\sigma_\textrm{bd}$. While this is supported by numerical simulations, as mentioned in the introduction, we also give a heuristic argument for it in appendix \ref{sec: chaos diffusion}. The single close approaches are combined below in \S \ref{sec:random walks}. The Delaunay variables are denoted by $(J_a,J_b,J_c,\theta_a,\theta_b,\theta_c)$ and are defined in, e.g., ref. \cite{Binney}.

Using a superscript or a subscript $s$ to denote variables pertaining to the outer binary, we have
\begin{align}
  L_z & = J_a^s \\
  L_x & = J_b^s\sin\theta_a^s\sin i_s \\
  L_y & = -J_b^s\cos\theta_a^s\sin i_s
\end{align}
This implies that the angular-momentum-conserving delta-function is
\begin{equation}
  \delta(\mathbf{J}-\mathbf{S}-\mathbf{L}) = \delta(J_x - S_x - J^s_b\sin\theta_a^s\sin i_s)\delta(J_y - S_y + J_b^s\cos\theta_a^s\sin i_s)\delta(J_z - S_z - J_a^s).
\end{equation}
This equation in turn implies that the angular-momentum integral is independent of $J_c^s$, modulo the integration domain boundaries, which are
\begin{align}
  & \abs{J_a^s} \leq J_b^s, \\ &
  0 \leq J_b^s \leq \min\set{A(E_\bin),J_c^s} \equiv \alpha.
\end{align}
(Please note that for the unbound case, $\alpha$ is defined simply as $A(E_\bin)$.)
The $\zhat$-axis integral gives
\begin{equation}
\begin{aligned}
    \sigma_\textrm{bd} & = \int_\bin \int\mathrm{d}J_c^s\mathrm{d}J_b^s\mathrm{d}\theta_a^s\mathrm{d}\theta_b^s\mathrm{d}\theta_c^s~\delta(\textrm{energy})\times \\ & \delta\left(J_x - S_x - J_b^s\sin\theta_a^s\sqrt{1 - \frac{(J_z-S_z)^2}{(J_b^s)^2}}\right)\delta\left(J_y - S_y + J_b^s\cos\theta_a^s\sqrt{1 - \frac{(J_z-S_z)^2}{(J_b^s)^2}}\right).
\end{aligned}
\end{equation}
As in ref. \cite{StoneLeigh2019} we perform a change of variables
\begin{equation}
  \left(\begin{array}{c}
          J_b^s \\
          \theta^s_a
        \end{array}\right) \mapsto \left(\begin{array}{c}
                                           z_1 \\
                                           z_2
                                         \end{array}\right) = \left(\begin{array}{c}
                                                                      \sin\theta_a^s\sqrt{(J_b^s)^2 - (J_z - S_z)^2} \\
                                                                      \cos\theta_a^s\sqrt{(J_b^s)^2 - (J_z - S_z)^2}
                                                                    \end{array}\right).
\end{equation}
The Jacobian of this transformation is
\begin{equation}
  \abs{\frac{\partial(\theta_a^s,J_b^s)}{\partial(z_1,z_2)}} = \frac{1}{\sqrt{z_1^2 + z_2^2 + (J_z - S_z)^2}}.
\end{equation}
Integrating over $z_1$, $z_2$ yields
\begin{equation}
  \sigma_\textrm{bd} = \int_{\bin ~\cap ~\set{\abs{\mathbf{J-S}} \leq \alpha}} \int \mathrm{d}J_c^s\mathrm{d}\theta_c^s\mathrm{d}\theta_b^s\frac{\delta\left(E - E_\bin + \frac{G^2M^2\mu_s^3}{2(J_c^s)^2}\right)}{\abs{\mathbf{J}-\mathbf{S}}}.
\end{equation}
The integral $\mathrm{d}\theta_b^s$ gives $2\pi$, while the integral $\mathrm{d}\theta_c^s$ -- over the mean anomaly -- gives a multiplicative factor of $\theta_{\max}$, which is the maximum mean anomaly the star may have and still stay in $\celtica$. Condition \eqref{eqn: close encounter condition} implies that $\theta_c^s = 0$ is in $\celtica$, while condition \eqref{eqn: hierarchy condition} implies that $\theta_c^s = \pi$ is no longer in $\celtica$. Thus, $\theta_{\max}$ restricts $\abs{\mathbf{r}_s}$ to $\abs{\mathbf{r}_s}\leq R$, such that the integration is carried out in $\celtica$.
The last lone-star integration, over $J_c^s$ may now be performed, to remove the last delta function, and give
\begin{equation}
  \sigma_{\textrm{bd}} = \frac{2\pi GM\mu_s^{3/2}}{\sqrt{8}}\int_{\bin ~\cap ~\set{\abs{\mathbf{J-S}} \leq \alpha}}\frac{\theta_{\max}}{\abs{\mathbf{J}-\mathbf{S}}\abs{E_0 - E_\bin}^{3/2}},
\end{equation}
where
\begin{equation}\label{eqn: theta max bound}
  \theta_{\max} = \arccos\left(\frac{a_s - R}{e_sa_s}\right) - \sqrt{\frac{2R}{a_s} - \frac{R^2}{a_s^2} - 1 + e_s^2}.
\end{equation}
One may perform the integration over the binary angles trivially, to give an additional factor of $(2\pi)^3$. This yields a cross-section
\begin{equation}\label{eqn:binary intermediate cross-section}
  \sigma_{\textrm{bd}} = \frac{(2\pi)^4 GM\mu_s^{3/2}}{\sqrt{8}}\int\frac{\theta_{\max}\mathrm{d}J_c\mathrm{d}J_b\mathrm{d}J_a}{\abs{\mathbf{J}-\mathbf{S}}\abs{E_0 - E_\bin}^{3/2}}.
\end{equation}
Taking the $\zhat$-axis in this integral to be along $\mathbf{J}$ implies that the integration domain is
\begin{equation}
  \begin{aligned}
    (J_a,J_b) & \in \set{(J_a,J_b): 0 \leq J^2 + J_b^2 - 2JJ_a \leq \alpha^2} \cap \set{\abs{J_a}\leq J_b} \\
    J_c & \in \set{E \leq E_\bin(J_c)\leq E_{\min}},
  \end{aligned}
\end{equation}
where $E_{\min}$ is some minimum cut-off on the binary energy. This form is the same as that of $\sigma_\textrm{ubd}$ of ref. \cite{StoneLeigh2019}, which implies that
\begin{equation}\label{eqn:binary cross-section}
  \sigma = \frac{(2\pi)^4 GM\mu_s^{3/2}}{\sqrt{8}}\int \frac{\theta_{\max}\mathrm{d}J_c\mathrm{d}J_b\mathrm{d}J_a}{\abs{\mathbf{J}-\mathbf{S}}\abs{E_0 - E_\bin}^{3/2}},
\end{equation}
where now the integration domain is
\begin{equation}\label{eqn:integration domain}
  \begin{aligned}
    (J_a,J_b) & \in \Omega \equiv \set{(J_a,J_b): 0 \leq J^2 + J_b^2 - 2JJ_a \leq \alpha^2} \cap \set{\abs{J_a}\leq J_b} \\
    J_c & \in \set{E_\bin(J_c)\leq \frac{E}{3}},
  \end{aligned}
\end{equation}
and $\theta_{\max}$ is defined by equation \eqref{eqn: theta max bound} for the bound case, and by equation \eqref{eqn: Stone and Leigh 2019} for the unbound case.
Equation \eqref{eqn:binary cross-section}, together with equations \eqref{eqn:integration domain}, specify the probability that the resultant binary has Delaunay actions $J_a,J_b,J_c$, given conserved quantities $E,\mathbf{J}$ -- their probability density function is simply the integrand in equation \eqref{eqn:binary cross-section}. It might be more useful to express equation \eqref{eqn:binary cross-section} in term of $E_\bin$, rather than $J_c$; the Jacobian for this transformation is simply $\propto E_\bin^{-3/2}$, which gives a distribution function for the outcome of one binary-single close approach,
\begin{equation}\label{eqn:f_bin}
  f_\bin(E_\bin,\mathbf{S}|E,\mathbf{J}) \propto \frac{E_\bin^{-3/2}}{\abs{\mathbf{J}-\mathbf{S}}\abs{E_0 - E_\bin}^{3/2}}\theta_{\max}(E_\bin,E_0,\mathbf{J}-\mathbf{S}).
\end{equation}
\end{widetext}

\section{Different Masses}
\label{sec:masses}
In fact, equation \eqref{eqn:f_bin} is proportional to the probability that a star $s \in \set{1,2,3}$ escapes, leaving a binary with energy $E_\bin$ and spin $\mathbf{S}$. What remains is the coefficient, which depends on the masses, as in equation \eqref{eqn:binary cross-section}. Therefore, the probability density that star $s$ is ejected after a close interaction, and that the remaining binary has energy $E_\bin$ and spin $\mathbf{S}$ is
\begin{widetext}
\begin{equation}\label{eqn:f_bin masses}
  f_\bin(E_\bin,\mathbf{S},s|E,\mathbf{J}) = N(s)(\mu_s\mu_\bin)^{3/2}\frac{m_\bin E_\bin^{-3/2}}{\abs{\mathbf{J}-\mathbf{S}}\abs{E_0 - E_\bin}^{3/2}}\theta_{\max}(E_\bin,E_0,\mathbf{J}-\mathbf{S}),
\end{equation}
where $N(s)$ is a normalisation constant, that depends on $s$ through the integration over \eqref{eqn:integration domain}.
To obtain this constant, one should change the integrands in $\sigma$ to ones that are independent of the masses, i.e. from $(J_a,J_b,J_c)$ in equation \eqref{eqn:binary cross-section} to semi-major axes, eccentricities and inclinations. The lone star's angular momentum and energy simply contribute a factor of $\mu_s^{-5/2}$, and the measure contributes an additional factor of $\mu_\bin^{3}m_\bin^{3/2}$. Thus, up to dimensionless quantities,
\begin{equation}\label{eqn:N(s)}
  N(s)\mu_s^{3/2} \propto \frac{\mu_\bin^3m_\bin^{3/2}}{\mu_s},
\end{equation}
where, now, the proportionality constant is independent of the identity of the ejected star (but may still depend on the total mass or on conserved quantities). One immediate prediction of equation \eqref{eqn:N(s)} is that, if one of the masses is much smaller than the other two, then the probability that each close interaction ends with the lighter one being shot out, rather than one of the others, dominates.
By dimensional analysis, the probability that mass $m$ escapes is therefore
\begin{equation}\label{eqn:ejected mass}
  P(m) \approx \frac{m_a^4 m_b^4}{\left(m_a+m_b\right){}^{5/2} \left[m_a^4 \left(\frac{m_b^4}{\left(m_a+m_b\right){}^{5/2}}+\frac{m^4}{\left(m_a+m\right){}^{5/2}}\right)+\frac{m^4 m_b^4}{\left(m_b+m\right){}^{5/2}}\right]}.
\end{equation}
\end{widetext}
We emphasise that equation \eqref{eqn:ejected mass} is an approximation, and for accurate results one should integrate equation \eqref{eqn:f_bin masses} over remnant binary energies or angular momenta or the marginal energy distribution, equation \eqref{eqn:f_E_bin marginal} below, over the allowed energies.\footnote{The reason it is not exact is that, when converting the integration over $E_\bin$ into an integration over $a_\bin$, one finds that there are now two parameters with dimension of length that can be used to re-scale $a_\bin$ for the purposes of dimensional analysis: $R$ and the original semi-major axis $a_0$. There is, therefore, an additional mass dependence hidden here, which equation \eqref{eqn:ejected mass} does not account for (but an integration of equation \eqref{eqn:f_E_bin marginal} over $E_\bin$ does).}

One can also compute the exchange cross-section, given by
\begin{equation}\label{eqn:exchange cross section masses}
\begin{aligned}
  \sigma(\textrm{Exchange}) & = \sigma_\textrm{tot} \times P(\textrm{Exchange}) \\ &
  = \frac{2\pi GM a_0}{v_0^2}P(\textrm{Exchange}),
\end{aligned}
\end{equation}
as the total cross-section is of course
\begin{equation}
  \sigma_\textrm{tot} = \frac{2\pi GM a_0}{v_0^2},
\end{equation}
where $a_0$ is the initial binary semi-major axis, and $v_0$ is the perturber's initial velocity (see, e.g. \citet{HeggieHut2003}).

\section{Detailed Balance}
\label{sec: detailed balance}
Let us try to obtain equation \eqref{eqn:binary cross-section} by an easier means. One might, for instance, assume that the triple is part of a globular cluster which is in thermal equilibrium, and which contains binaries, single stars, and triples. Of course, the desired cross-section $\sigma$ does not depend on whether there exists such a cluster, and on whether this hypothetical cluster is indeed in thermal equilibrium. Making those auxiliary assumptions would simplify the calculation, because then we could use the principle of \emph{detailed balance} (cf. refs. \cite{Heggie1975,HeggieHut1993,HeggieHut2003}).

If the cluster is in thermal equilibrium, then the numbers of binaries with energy $E_\bin$ and stars with energy $E_s$ must be constant. This number density is proportional (in the canonical ensemble, neglecting collisions and stellar evolution) to $\exp(-\beta^* \ham)$, where $\beta^* = 1/(k_B T)$. From this, one may deduce that the rates at which encounters between stars and binaries transfer these systems from one energy to another, and those of the reverse process, have to be equal. This is the principle of detailed balance.

Let $\Gamma(E_\bin,\mathbf{S},\mathbf{p} \to E_0,\mathbf{J})\mathrm{d}E_0\mathrm{d}\mathbf{J}\mathrm{d}E_\bin\mathrm{d}J_a\mathrm{d}J_b$ be the differential rate at which binaries and single stars combine to form bound triples with energy $E_0$ and total angular momentum $\mathbf{J}$, and likewise let $\Gamma(E_0,\mathbf{J} \to E_\bin,\mathbf{S})\mathrm{d}E_0\mathrm{d}\mathbf{J}\mathrm{d}E_\bin\mathrm{d}J_a\mathrm{d}J_b$ be the rate of disintegration of such triples. Let $n_\textrm{triple}(E_0,\mathbf{J})$ denote the number density of such triples, and let $n_\bin(E_\bin,\mathbf{S})$ and $n(\mathbf{p}_3)$ denote the phase-space densities of binaries and single stars, respectively. Detailed balance amounts to the requirement that the number of bound triples remain constants, i.e. that the rate of the forward and backward reactions, weighed by the relevant densities, cancel each other out. Symbolically, in barycentric co-ordinates
\begin{widetext}
\begin{equation}\label{eqn:detailed balance}
  n_\textrm{triple}(E_0,\mathbf{J})\Gamma(E_0,\mathbf{J} \to E_\bin,\mathbf{S}) = \int \mathrm{d}^3\mathbf{R}_\textrm{cm}\mathrm{d}^3\mathbf{P}\mathrm{d}^3\mathbf{p}\delta\left(\mathbf{R}_\textrm{cm}\right)\delta\left(\mathbf{p}_\bin + \mathbf{p}_3\right) \times n(\mathbf{p}_3)n_\bin(E_\bin,\mathbf{S})\Gamma(E_\bin,\mathbf{S},\mathbf{p} \to E_0,\mathbf{J}),
\end{equation}
where $\mathbf{p}$ is the momentum of the relative motion between the third star and the binary centre-of-mass. (See, e.g., ref. \cite{HeggieHut1993} for a derivation of a similar expression for the reaction rate.)

The advantage of equation \eqref{eqn:detailed balance} is that the rate of formation of bound triples may be simpler to compute; one could approximate it as
\begin{equation}\label{eqn:rate of formation of triples with angular momentum}
\begin{aligned}
  n_\bin(E_\bin,\mathbf{S})\Gamma(E_\bin,\mathbf{S} \to E_0,\mathbf{J}) & = \int\mathrm{d}^3\mathbf{p}~n(\mathbf{p}_3)n_\bin(E_\bin,\mathbf{S})\Gamma(E_\bin,\mathbf{S},\mathbf{p} \to E_0,\mathbf{J}) \\ &
  = \int \mathrm{d}^3\mathbf{p} \iint_{D(v)} \mathrm{d}^2\mathbf{b}~ v ~\delta\left(E_0 - E_\bin - E_s\right)\delta\left(\mathbf{J} - \mathbf{S} -\mu_s bv\phihat \right) n(\mathbf{p}_3)n_\bin(E_\bin,\mathbf{S})
\end{aligned}
\end{equation}
\end{widetext}
where $D(v)$ is a disc in the $\xhat$-$\yhat$ plane at infinity, with radius $a_\bin \sqrt{1 + \frac{GM}{\mu_s v^2a_\bin}}$ (see, e.g. ref. \cite{HeggieHut1993}).\footnote{We remark that one cannot assume that $\mu_sv^2$ is much smaller than $E_\bin$; if we did, then thermal equilibrium would be precluded, as interactions between hard binaries and single stars are known to be a source of heat for globular clusters (see, e.g, ref. \cite{HeggieHut2003}).} The direction of the $\zhat$-axis is chosen so that it points along $\mathbf{v}$, (assumed not to be away from the binary, for then there wouldn't be any encounter). The integral over $\mathbf{b}$, including the angular momentum delta-function gives
\begin{equation}
  \delta(\theta_{\mathbf{J}-\mathbf{S}} - \pi/2)\times \begin{cases}
    \frac{1}{\mu_s^2v^2\abs{\mathbf{J}-\mathbf{S}}}, & \mbox{if } (\mathbf{J}-\mathbf{S})^2 < A^2\\
    0, & \mbox{otherwise}.
  \end{cases}.
\end{equation}
The remaining delta-function should not really be there -- it is just a mathematical artefact, which came from the way we defined the axes, so it is safe to omit it. Alternatively, one could justify its omission by integrating over all possible orientations of the axes: this delta function then picks out the orientation where the $\zhat$ axis is perpendicular to $\mathbf{J}-\mathbf{S}$; such an integration is in turn justified by the fact that a choice of axis is meaningful only for the right-hand side of equation \eqref{eqn:detailed balance}, and not for the left-hand side. The factor of $\abs{\mathbf{J}-\mathbf{S}}$ in the denominator comes from expressing the angular momentum delta function in spherical co-ordinates: this turns the 3-dimensional Dirac delta function $\delta\left(\mathbf{J} - \mathbf{S} - \mathbf{L}\right)$ (recall that $\mu_s bv\phihat = \mathbf{L}$) into a product of three 1D delta functions, one for the magnitudes of $\mathbf{J}-\mathbf{S}$ and $\mathbf{L}$, one for one angle, and one for the other angle, divided by the appropriate Jacobian $\abs{\mathbf{L}}^2\sin\theta_{\mathbf{J}-\mathbf{S}}$. Since the other delta function sets $\theta=\pi/2$, we are left with $\abs{\mathbf{L}}^2$ in the denominator, and in the numerator: two angular delta functions multiplied by $\delta\left(\abs{\mathbf{J} - \mathbf{S}} - \abs{\mathbf{L}}\right)$. We now perform the integral $\mathrm{d}^2\mathbf{b}$ in polar co-ordinates $b$ and $\varphi$, by writing $\abs{\mathbf{L}}= \mu_s bv$, and changing variables from $\abs{\mathbf{L}}$ to $b$. As $\mathrm{d}^2\mathbf{b} = b\mathrm{d}b\mathrm{d}\varphi$, one $b$ cancels one power of $\abs{\mathbf{L}}$ from the denominator, and the integral $\mathrm{d}b$ removes the absolute value delta function, replacing the other $\abs{\mathbf{L}}$ with $\abs{\mathbf{J}-\mathbf{S}}$. The integration $\mathrm{d}\varphi$ removes one of the angular delta functions, and the second one is removed as explained above.

The density $n(\mathbf{p}_3)$ is a Maxwell-Boltzmann distribution, which is proportional to $\rho m_3^{-3/2}\exp(-\beta^* E_s)$.
After performing the integral over $\mathbf{p} = \mu_s \mathbf{v}$, keeping track of the energy-conserving delta-function, one has $\Gamma(E_\bin,\mathbf{S}\to E_0,\mathbf{J}) \propto \frac{\mu_s^3m_3^{-3/2}}{\mu_s^3\abs{\mathbf{J}-\mathbf{S}}}e^{\beta^*(E_\bin - E_0)}$. \citet{Heggie1975} gives a formula for the density of binaries with energy $E_\bin$, eccentricity $e$ and inclination $i$ \citep[equation 2.12]{Heggie1975}, from which the phase-space density $n_\bin(E_\bin, \mathbf{S})$ is determined to be $\propto \rho^2(m_1m_2)^{-3/2}\mu_\bin^{3/2}m_\bin e^{-\beta^* E_\bin} E_\bin^{-3/2}$. Hence, by detailed balance
\begin{equation}\label{eqn: disintegration rate detailed balance}
  \Gamma(E_0,\mathbf{J} \to E_\bin,\mathbf{S}) \sim \frac{\rho^3e^{-\beta^* E_0}(m_1m_2m_3)^{-3/2}m_\bin\mu_\bin^{3/2}}{n_{\rm triple}(E_0,\mathbf{J})E_\bin^{3/2}\abs{\mathbf{J}-\mathbf{S}}},
\end{equation}
provided that the condition $(\mathbf{J}-\mathbf{S})^2 < \alpha^2$ obtains.

How is this rate related to $f_\bin(E_\bin,\mathbf{S}|E_0,\mathbf{J})$? By definition, it is the number of disintegrations per unit time, i.e., it is the number of triples in which star $s$ is between $\theta_c^s$ and $\theta_c^s + \Omega_c^s\mathrm{d}t$, divided by $\mathrm{d}t$, where $\theta_c^s \in [0,\theta_{\max}]$. That is,
\begin{equation}
  \Gamma(E_0,\mathbf{J} \to E_\bin,\mathbf{S}) \propto \Omega_c^s \frac{\mathrm{d}f_\bin(E_\bin,\mathbf{S}|E_0,\mathbf{J})}{\mathrm{d}\theta_c^s},
\end{equation}
on the one hand, and on the other hand we have equation \eqref{eqn: disintegration rate detailed balance}. Together these imply that upon division by the lone star's orbital frequency $\Omega_c^s \propto \abs{E_s}^{3/2}/(M\mu_s^{3/2})$ and integration over $\theta_c^s$,
\begin{equation}
  f_\bin(E_\bin,\mathbf{S}|E_0,\mathbf{J}) \propto \begin{cases}
                                               \frac{m_\bin\theta_{\max}\abs{\mathbf{J}-\mathbf{S}}^{-1}}{\abs{E_s}^{3/2}\abs{E_\bin}^{3/2}}, & \mbox{if } \abs{\mathbf{J}-\mathbf{S}} \leq \alpha \\
                                               0, & \mbox{otherwise}.
                                             \end{cases}
\end{equation}
up to a function symmetric in all the particle masses, as in equation \eqref{eqn:f_bin}. (The expressions $\mu_s\mu_\bin $, $m_1m_2m_3$ and $M$ are all such symmetric functions.)

\section{Marginal Energy Distribution}
\label{sec:marginal energy distribution}
Let us compute the marginal energy distribution
\begin{equation}
  f_\bin(E_\bin|E,\mathbf{J}) = \int_{\Omega} \mathrm{d}J_a\mathrm{d}J_b f_\bin(E_\bin,\mathbf{S}|E,\mathbf{J}).
\end{equation}
For this purpose, let
\begin{equation}\label{eqn: scri}
  \scri(\alpha) = \int_{\Omega} \frac{\mathrm{d}J_b\mathrm{d}J_a}{\abs{\mathbf{J} - \mathbf{S}}}\frac{\theta_{\max}}{\theta_{\max}(e_s = 1)};
\end{equation}
$\scri$ is the integral one must evaluate. The calculation is performed in appendix \ref{appendix: angular momentum integral}, and the outcome is that $\scri$ is well-approximated by a power-law, proportional to $E_\bin^{-1/2}$ for low $J$, but to $E_\bin^{-1}$ for large values of $J$.

A consequence of \S \ref{subsec: Celtica in angle-action variables} (inequality \eqref{eqn: energy restriction bound}) is that for $E_\bin > E_0$ -- i.e. in the bound case -- the final binary energy is constrained to lie close to the total energy. In this neighbourhood, $\theta_{\max} \sim \abs{E - E_\bin}^{3/2}$, which cancels the existing $\abs{E - E_\bin}^{-3/2}$. Outside this region, $\theta_{\max}$ is approximately constant; thus, one may remove its eccentricity dependence by approximating $e_s \approx 1$, writing
\begin{widetext}
\begin{equation}
  \theta_{\max} \approx \theta_{ap}(E_\bin) \equiv \begin{cases}
                                                     \arccos\left(1-\frac{R}{a_s}\right) - \sqrt{2\frac{R}{a_s} - \frac{R^2}{a_s^2}}, & \mbox{bound case} \\
                                                     \sqrt{2\frac{R}{a_s} + \frac{R^2}{a_s^2}} - \arccosh\left(1+\frac{R}{a_s}\right), & \mbox{unbound}.
                                                   \end{cases},
\end{equation}

This implies that the marginal energy distribution is
\begin{equation}
\label{eqn:f_E_bin marginal}
  f_\bin(E_\bin|E,\mathbf{J}) \propto m_\bin\begin{cases}
                                  \frac{\scri(\alpha(E_\bin))\theta_{ap}(E_\bin)}{\abs{E_\bin}^{3/2}\abs{E - E_\bin}^{3/2}}, & \mbox{if } E_\bin > E, ~\abs{E - E_\bin} \leq \frac{Gm_\bin\mu_s}{\eta R} \\
                                  \frac{\scri(A(E_\bin))\theta_{ap}(E_\bin)}{\abs{E_\bin}^{3/2}\abs{E - E_\bin}^{3/2}}, & E_\bin \leq E.
                                \end{cases}
\end{equation}
\end{widetext}


\section{A Random-Walk Description}
\label{sec:random walks}
We have now reached the point where we may introduce a random-walk description of the evolution between consecutive close triple approaches. Suppose that during each close approach, the constants of motion $(E,\mathbf{J})$ might change by some amount, according to some probability distribution, which depends on the state of the system at the beginning of the close approach (i.e. at the end of the previous one), because of some additional astrophysical process. Denote this distribution by $f_c(E^k,\mathbf{J}^k|E^{k-1}_\bin,E^{k-1},\mathbf{S}^{k-1},\mathbf{J}^{k-1})$, where $E^j$, \emph{etc.} denote quantities at the end of the $j$-th close approach. A definition of such an additional astrophysical process one would like to incorporate in one's study of binary-single encounters amounts, therefore, to providing $f_c$. Otherwise, $f_c = \delta(E^k - E^{k-1})\times \delta\left(\mathbf{J}^k - \mathbf{J}^{k-1}\right)$, by default.

The total energy $E$, the total angular momentum $\mathbf{J}$, the spin $\mathbf{S}$ and the binary energy $E_\bin$ thus perform a random walk, where the probabilities for the $j$-th value are dictated by the values of these quantities at the $j-1$-th step -- this random walk has a one-step memory. This process may describe, for example, a tidal interaction between two stars during the encounter (see \S \ref{sec:numerical simulations} below).

The beauty of this description is that now one can use it to find the ultimate binary parameter distribution $P(E_\bin,\mathbf{S})$, when the single star leaves, never to return. 
Let $x = (E_\bin,E,\mathbf{S},\mathbf{J})$, and let $h(x|x')$ denote the probability of the walker (i.e. the binary + single) moving from $x'$ to $x$ at one step -- that is, the probability that it started the close approach at $x'$ and left it at $x$. Explicitly
\begin{equation}
  h(x|x') = f_\bin(E_\bin,\mathbf{S}|E,\mathbf{J})f_c(E,\mathbf{J}|E',E_\bin',\mathbf{S}',\mathbf{J}').
\end{equation}
The ultimate reason we spent so much effort above computing $f_\bin$ for the bound three-body problem is precisely so that we would know what $h(x|x')$ looks like. The mixing hypothesis ensures that the functional form of the way $h$ depends on the $E_\bin, \mathbf{S}$ components of $x$ is only through $f_\bin$ as calculated in \S \ref{sec: phase space integration}. 
In particular, it also allows us to account spatial cut-offs; e.g. in a dense cluster environment an ejected, but still bound third star, which would have otherwise eventually fallen back into $\celtica$, would now be met with an external perturbation by other stars, if its separation became comparable to the distance between stars in the cluster. In other words, the environment could potentially dictate an effective binding energy limit which would be different from the clean case of an isolated interacting triple. Our model can easily incorporate this aspect.

Let us also introduce the following linear, integral operators, acting on a function $\varphi(x)$:
\begin{widetext}
\begin{align}
  & (W_{\textrm{lim}}\varphi) (x) = \int_{\set{E_\bin' \geq E'}} \mathrm{d}x' h(x|x')\varphi(x') \\ &
  (W_{\textrm{unlim}}\varphi) (x) = \int_{\set{E_\bin' < E'}}\mathrm{d}x' h(x|x')\varphi(x').
\end{align}
The first describes an encounter that ends with the third star bound, and the second -- the final encounter. Suppose we start with initial probability
\begin{equation}\label{eqn:initial probability}
  p_{i}(x) = N f_\bin(E_\bin,\mathbf{S}|E_0,\mathbf{J}_0)\delta(E-E_0)\delta(\mathbf{J}-\mathbf{J}_0),
\end{equation}
where $E_0$ is the initial total energy and $\mathbf{J}_0$ is the initial total angular momentum. Now,
\begin{align}
  W_{\textrm{lim}}(p_i) & = Nf_\bin(E_\bin|E,\mathbf{J})\int \mathrm{d}E'\mathrm{d}^3\mathbf{J}'\mathrm{d}E_\bin'\mathrm{d}^2\mathbf{S}'~ f_\bin(E_\bin',\mathbf{S}'|E',\mathbf{J}')\delta(E'-E_0)\delta(\mathbf{J}'-\mathbf{J}_0)f_c(E,\mathbf{J}|x') \\ &
  = Nf_\bin(E_\bin|E,\mathbf{J})\int \mathrm{d}E_\bin'\mathrm{d}^2\mathbf{S}'~f_\bin(E_\bin',\mathbf{S}'|E_0,\mathbf{J}_0)f_c(E,\mathbf{J}|E_0,E_\bin',\mathbf{S}',\mathbf{J}_0).
\end{align}
This means that the only dependence on $E_\bin,\mathbf{S}$ is outside the integral, inside an $f_\bin$ -- just as in equation \eqref{eqn:initial probability}. In-so-far-as the binary actions are concerned, the action of $W_\textrm{lim}$ does not alter the functional form of the probability distribution.

Now suppose that $f_c$ may be expanded as a sum of changes in energy and angular momentum, relative to $E_0,\mathbf{J}_0$, whose probabilities depend on the previous round:
\begin{equation}
  f_c(E,\mathbf{J}|E',E_\bin',\mathbf{S}',\mathbf{J}') = \int \mathrm{d}\lambda\mathrm{d}^3\chi ~\delta(E- (E_0 - \lambda))\delta(\mathbf{J} - (\mathbf{J}_0 - \chi))p_E(\lambda,\chi|E_\bin',\mathbf{S}',E_0,\mathbf{J}_0);
\end{equation}
(this is the law of total probability in disguise) if this is the case, then
\begin{equation}\label{eqn: w_lim of f_bin}
  W_{\textrm{lim}}(p_i) = Nf_\bin(E_\bin,\mathbf{S}|E,\mathbf{J})\int \mathrm{d}\lambda\mathrm{d}^3\chi ~\tilde{p}_E(\lambda,\chi;E_0,\mathbf{J}_0)\delta(E-(E_0 - \lambda))\delta(\mathbf{J} - (\mathbf{J}_0 - \chi)),
\end{equation}
\end{widetext}
where we have defined
\begin{equation}
  \tilde{p}_E(\lambda,\chi;E_0,\mathbf{J}_0) = \int \mathrm{d}E_\bin'\mathrm{d}^2\mathbf{S}' ~p_E(\lambda,\chi|E_\bin',\mathbf{S}',E_0,\mathbf{J}_0).
\end{equation}
Equation \eqref{eqn: w_lim of f_bin} implies that one encounter, if the initial probability distribution was some constant times $f_\bin$ and a total-energy-total-angular-momentum delta-function, turns this form into a sum of terms of similar structure. This fact helps us to find the final distribution of binary energies and spins, when tidal interactions are taken into account, as we do below.

Using the particularly special form of the action of $W_\textrm{lim}$ on $p_i$ (the action of $W_\textrm{unlim}$ is quite similar), we may determine the final distribution in terms of the initial total energy and angular momentum. This is done in a perturbative manner, assuming that the probability of a non-zero change in these quantities is small.\footnote{Here we sum up all orders, so that this assumption is innocuous in-so-far-as the general model described in this section is concerned. In \S \ref{sec:numerical simulations} Below we truncate the series at linear order.} Let $P_n(x|x_0)$ be the probability that the full close interaction ends after exactly $n$ steps, at $x$, given that it started out initially at $x_0$. By the properties of random walks (see, e.g. \citet{BarryHughes1995}),
\begin{equation}
  P_n(x) = \int \mathrm{d}x_1\ldots\mathrm{d}x_{n-1} h(x|x_{n-1})\cdot\ldots\cdot h(x_1|x_0)p_i(x_0),
\end{equation}
integrated over
\begin{equation}
  \set{E^k_\bin \geq E^k}_{k=1}^{n-1}.
\end{equation}
$P_n$ is therefore $W_{\textrm{unlim}}$ acting once after $n-1$ actions of $W_\textrm{lim}$, on $p_i$. The final probability is
\begin{equation}\label{eqn:random walk final probability}
  P(x) = \sum_{n=0}^{\infty}P_n(x).
\end{equation}
Equation \eqref{eqn:random walk final probability} is the ultimate distribution of binary parameters, after a complete binary-single encounter, incorporating the physical process described by $f_c$, in addition to the classical three-body dynamics. We apply this formalism to include the effects of tides and collisions in \S \ref{sec:numerical simulations}. Equation \eqref{eqn:random walk final probability} implies that if the conserved quantities don't change, then $P(x)$ is just $f_\bin$, and the entire process is rendered memory-less.
We now move on to compare the theoretical predictions made in this paper with results of numerical simulations.

\section{Comparison With Simulations -- Marginal Distributions}
\label{sec:comparison with simulations -- marginal distributions}
Let us start by comparing some marginal distributions of $f_\bin$ to numerical simulations, before moving on to test the full random-walk model in \S \ref{sec:numerical simulations}. All numerical integrations in this paper were done using {\small MATLAB}'s \verb"integral" functions.

\begin{figure*}
  \centering
  \includegraphics[width=0.45\textwidth]{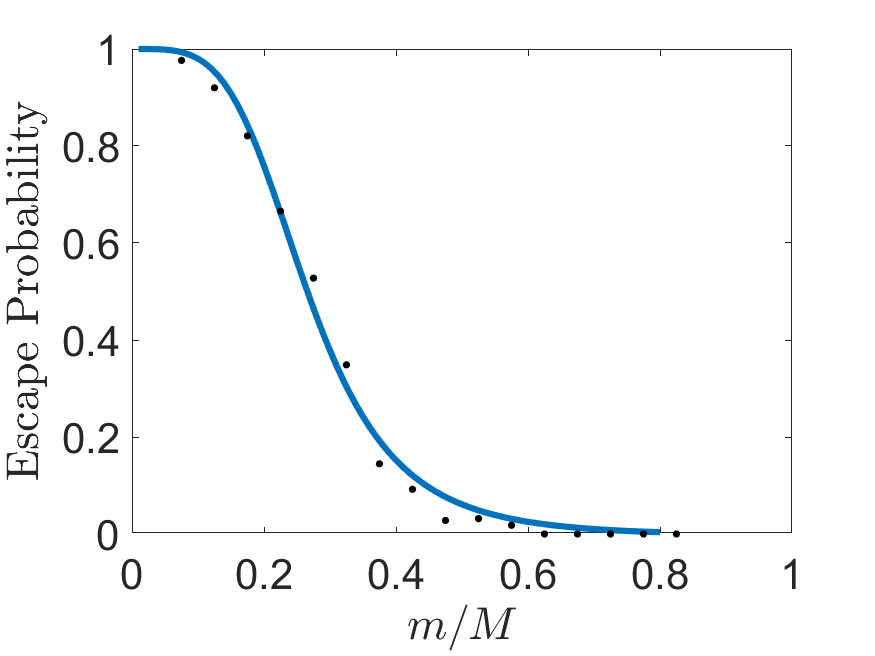}
  \includegraphics[width=0.45\textwidth]{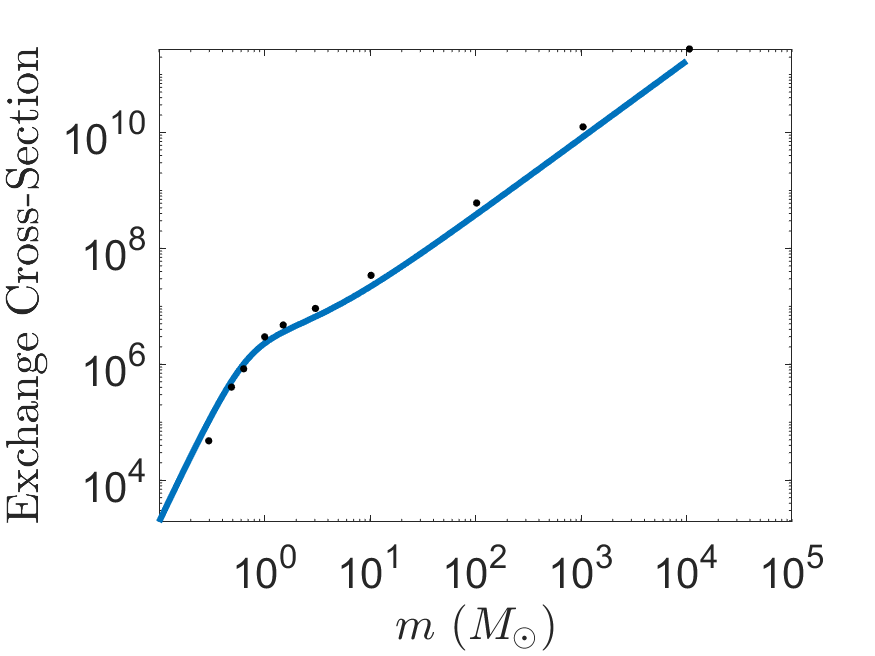}
  \caption{Left: The probability that a star of mass $m$ escapes, where $m_a = m_b = 1 ~M_\odot$, compared with data from ref. \cite{Saslawetal1974}. Right: The exchange cross-section, in units of the geometric cross-section, $\pi a_0^2$, for binary masses equal to $1 M_\odot$, and third star mass $m$, which arrives with initial velocity $v = 0.001v_\textrm{orb}$, compared with data from ref. \cite{Hills1992}.}\label{fig:masses}
\end{figure*}

We start with testing the ejected mass probability, which is given by equation \eqref{eqn:ejected mass}, to two simulations, by \citet{Saslawetal1974} and by \citet{Hills1992} in figure \ref{fig:masses}. Please bear in mind that when the perturbing star's mass is much larger than the initial binary members' masses, $a_0$ in equation \eqref{eqn:exchange cross section masses} needs to be modified by another multiplicative factor of $[M/(m_a+m_b)]^{1/3}$, due to an increased effective total cross-section -- this is what we show on the right panel of figure \ref{fig:masses}.

\begin{figure*}
  \centering
  \includegraphics[width=0.9\textwidth]{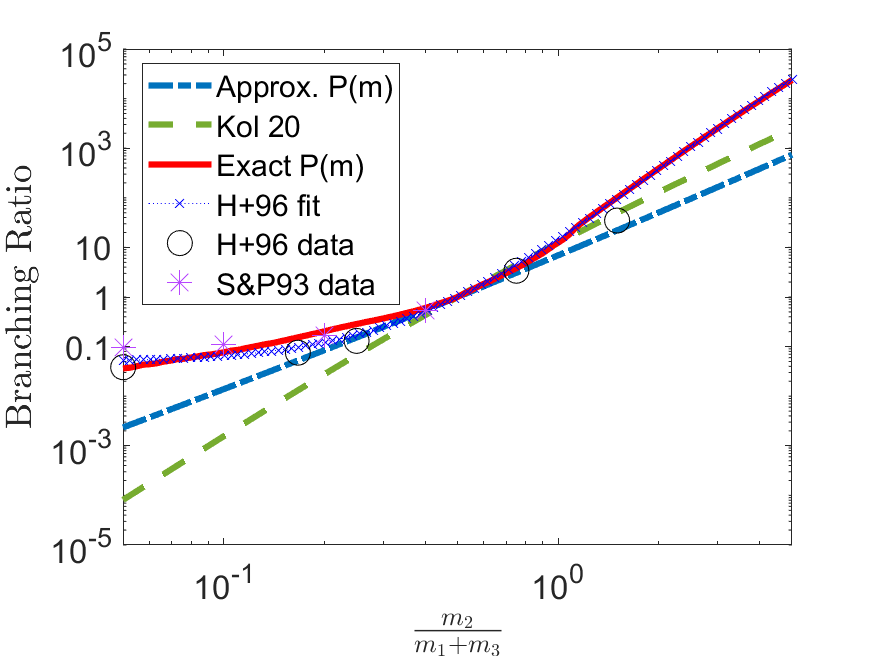}
  \caption{The predictions of equation \eqref{eqn:ejected mass} (sea blue, dash-dotted), equation \eqref{eqn:f_E_bin marginal} (red), integrated over the relevant range, and equation A.16 of \citet{Kol2020} (green, dashed), compared with the semi-analytical fit of \citet{Heggieetal1996} (blue, dotted with crosses) and numerical simulation data from \citet{Heggieetal1996} (black circles) and \citet{SigurdssonPhinney1993} (purple asterisks). The initial binary masses are $m_1 = 1 ~M_\odot$, and $m_2$, and the in-coming star's mass is $m_3 = m_1$. Its velocity is $v_0 = 0.1v_c$ (where $v_c$ is defined in the caption of figure \ref{fig:SP_a}) to ensure that the binary is hard. The $y$-axis shows the branching ratio of the cross-section for ejecting $m_1$ relative to the cross-section for ejecting $m_2$. There is excellent agreement with equation \eqref{eqn:f_E_bin marginal}.}\label{fig:Heggie masses}
\end{figure*}

We also test the predictions of our model by comparing them to the simulations results of ref. \cite{Heggieetal1996}, who calculated exchange cross-sections for a wide range of masses. In figure \ref{fig:Heggie masses}, we plot the predictions obtained by integrating equation \eqref{eqn:f_E_bin marginal} over the allowed range, as well as those of the approximate equation \eqref{eqn:ejected mass}, and those of appendix A of ref. \cite{Kol2020}, for the following situation: the initial binary consists of masses $m_1 = 1~ M_\odot$, and $m_2$, and the in-coming star has mass $m_3 = m_1$; we plot the branching ration, defined by $BR = \frac{\sigma_{\textrm{exch}(1)}}{\sigma_{\textrm{exch}}(2)}$, i.e. by the ratio of the exchange cross-section for ejecting star $1$, and that of ejecting star $2$.\footnote{The cross-sections were found by integrating the ejection probabilities \emph{given} angular momentum $J$ (and energy $E$) over the allowed values of the impact parameter $b$, using the law of total probability: each values of $b$ determines the angular momentum $J$, and thence the ejection probabilities, which are the integrated to yield the cross-section.} Ref. \cite{Heggieetal1996} provides an analytical fit, which is also plotted, as well as data from ref. \cite{SigurdssonPhinney1993}, for comparison. This fit is based on the entirety of the numerical simulations in ref. \cite{Heggieetal1996}, which cover an extensive range of mass ratios. The data from ref. \cite{Heggieetal1996} in this figure are for resonant cross-sections, while the fit is, to our understanding, for the total one, which is dominated by the resonant cross-section everywhere, especially for large mass-ratios.

As one is concerned with exchange cross-sections, they decay to zero at sufficiently large impact parameters. The impact parameter serves only to determine the total angular momentum, and as the exchange cross-section tends to zero as $J \to \infty$, one can integrate equivalently over $\mathbf{J}$. There is a natural cut-off $J_*$, which is the maximum angular momentum for which $\scri$ does not vanish (for any $E_\bin$), minimised over all ejected masses. Above $J_*$, there is no configuration in which the triple could have been in a non-hierarchical phase before separating (cf. \S \ref{subsec:in-spiral cross-section} below). The reader should bear in mind that if $m_2$ is considerably larger than $m_1$, then $BR \gg 1$, since there is a much-higher probability to eject the light particle, and likewise, for $m_2 \ll m_1, m_3$, $BR$ should decay to zero. All theoretical predictions plotted in figure \ref{fig:Heggie masses} satisfy these limits, but only the exact prediction of equation \eqref{eqn:f_E_bin marginal} meshes well with the data and with the semi-analytical fit.

Next, we move on to show the semi-major axis distribution. Ref. \cite{StoneLeigh2019} already obtained a good agreement between numerical simulations and the unbound cross-section, which has a similar form to equation \eqref{eqn:binary cross-section}. To check whether the bound one is also correct, we compare $f_\bin$ to the numerical results of \citet{SigurdssonPhinney1993}, who imposed an energy cut-off on the ejected star. There, as mentioned above, even some encounters with the lone star ejected with negative energy were deemed to be concluded, if its semi-major axis was large enough. This was meant to mimic the environmental effect of the globular cluster, where the triple resides. Clearly, once a single star is sufficiently far from the binary, it feels the cluster's potential more strongly and ceases to be bound to the binary, even if its energy is negative. This environmental cut-off implies that one has to use both $\sigma_\textrm{bd}$ and $\sigma_\textrm{ubd}$ to match the numerical results of ref. \cite{SigurdssonPhinney1993}.

\begin{figure*}
  \centering
  \includegraphics[width=0.9\textwidth]{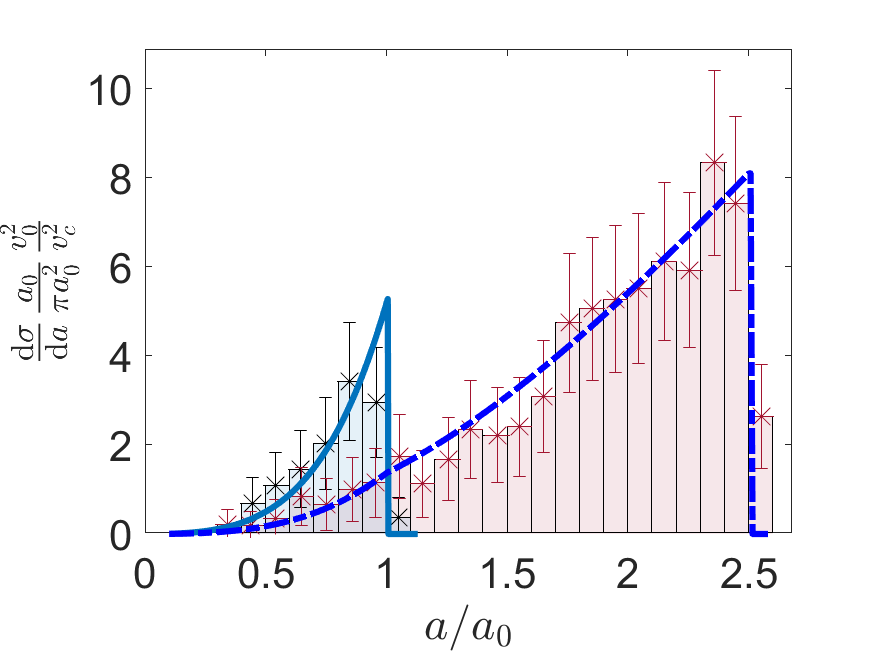}
  \caption{A comparison between equation \eqref{eqn:f_E_bin marginal} and the numerical simulations of \citet[figure 2b]{SigurdssonPhinney1993}. The initial semi-major axis is $a_0 = 0.1$ AU and the initial eccentricity is $e_0 = 0$; the initial binary masses are $m_1 = 1.4~M_\odot$, $m_2 = 0.56~M_\odot$, and the incoming third star's mass is $m_3 = 1.4~M_\odot$. Its initial velocity $v_0$ is uniformly distributed between $0.05v_c$ and $0.15v_c$, where $v_c = \sqrt{GM\mu_\bin/(m_3a_0)}$. The impact parameter is uniformly distributed in a disc $D(v_0)$ at infinity, whose radius is $b_{\max}(v_0) = a_0\left(4v_c/v_0 + 0.6(1+e_0)\right)$. All these parameters were chosen to match those of ref. \cite{SigurdssonPhinney1993}. We use $\beta = 1.5$ but the results are insensitive to $\beta$. The straight line and the data marked by blue bars pertains to the exchange $(1,2) + (3) \to (1) + (2,3)$, while the dashed-dotted line and the data marked by purple bars are for the process $(1,2) + (3) \to (1,3) + (2)$, where the lighter star is ejected. Error-bars correspond to $3\sigma$ statistical (Poisson) errors, arising from a total integration number of $4000$ \cite[figure 9]{SigurdssonPhinney1993}. As expected by \citet{SigurdssonPhinney1993}, the cut-off at large values of $a$ is faster than exponential, occurring almost instantaneously. Data for fly-bys is not shown here, as in \citet{SigurdssonPhinney1993} it is dominated by adiabatic fly-bys, and the effect of resonant scattering is hard to disentangle from it.\footnote{Observe that as the masses are different, the factor of $m_\bin$ in equation \eqref{eqn:binary cross-section} must be incorporated as well. Thus the normalisation of equation \eqref{eqn:f_E_bin marginal} is the sum, over all three possible final states, of the integrals of $f_\bin(E_\bin|E,\mathbf{J})$ over $E_\bin$. Also note that the graphs shown here are normalised such that the cross-sections (integrated over $a$) match the values of \citet[table 3B]{SigurdssonPhinney1993}. This is necessary as the total cross-section we calculate only takes close encounters into account, while the total cross-section of \citet{SigurdssonPhinney1993} also includes weak interactions.}}\label{fig:SP_a}
\end{figure*}

We do so by modifying the marginal energy distribution in equation \eqref{eqn:f_E_bin marginal} to account for the external cut-off criterion. Explicitly, \citet{SigurdssonPhinney1993} took a cut-off of $a_s(1+e_s) = 960a_0$, where $a_0$ is the initial semi-major axis of the binary. When the apoapsis was larger than this value, they considered the third body to be unbound from the binary. This may be incorporated into $f_\bin$ simply by modifying the apoapsis criterion in equation \eqref{eqn: close encounter condition} accordingly. The result is compared with their simulation results in figure \ref{fig:SP_a}.

As one can tell from figures \ref{fig:masses}, \ref{fig:Heggie masses} and \ref{fig:SP_a}, the function $f_\bin$ calculated above agrees well with simulation results. We now proceed to test the random walk model in more detail in the next section.

\section{Comparison With Simulations -- Tides}
\label{sec:numerical simulations}
To test the random-walk model described in \S \ref{sec:random walks}, we assume that dissipation is caused by tides, and we compare with the extensive $3$-body simulations conducted by \citet{Samsingetal2017}, which include both tidal forces and relativistic corrections. They investigated many cases, and we choose the equal mass case $m = 1.2M_{\odot}$; the initial binary is comprised of a white dwarf and a compact object (i.e. point particle), and the incoming lone star is also a point particle. The white dwarf's radius is $r_* = 0.006R_\odot$. The initial orbital separation is $a_0$, and the initial speed of the third body is $v_0 = 10~\textrm{km s}^{-1}$. Its origin is sampled uniformly from a disc $D(v_0)$ whose radius is \citep{Samsingetal2014}
\begin{equation}
  b_{\max} = \frac{a_0}{2}\sqrt{1+\frac{4GM}{a_0v_0^2}}.
\end{equation}

Before proceeding, let us note that equation \eqref{eqn: w_lim of f_bin}, together with the form of $f_\bin$, suffices to explain the main features of numerical simulations: inserting equation \eqref{eqn: w_lim of f_bin} into equation \eqref{eqn:random walk final probability}, implies that
\begin{equation}\label{eqn:encounters final probability}
  P(x) = f_\bin(E_\bin,\mathbf{S}|E,\mathbf{J})\int\mathrm{d}\tilde{\lambda}\tilde{p}(\tilde{\lambda}|E_0)\delta(E-(E_0 - \tilde{\lambda})),
\end{equation}
where the upper $\sim$ signs indicate that the energy-shifts and their associated probabilities may be different from those obtained in a single action of $W_\textrm{lim}$, but the structure is still the same. The final energy is `traced out', while the initial total energy is fixed.

Below we start by describing the tidal model we adopt in this paper, then we compute $P(x)$ perturbatively, and then compute the cross-section for a collision and a tidal in-spiral. The latter event is an in-spiral of two stars into one another due to orbital energy loss to tides. We back our analysis up by comparing its results to the numerical simulations of ref. \cite{Samsingetal2017}.

\subsection{Tidal Model}

We adopt the tidal model of ref. \cite{PressTeukolsky1977}. We further deem any energy that goes into the tidal oscillations of the white dwarf as lost from the system (that is, the time-scale on which it might return to orbital energy is much larger than the relevant dynamical time-scales), and we approximate the total angular momentum as fixed.\footnote{This is justified by the fact that the amount of angular momentum that goes into the tidal excitations is $\sim \sqrt{\frac{r_*^3}{Gm}}\Delta E$ \citep{Kochanek1992}, which is very small in comparison with the initial orbital angular momentum of the binary, $\mu_\bin \sqrt{Gm_\bin a_0}$.} We take a tidal dissipation event in a single close approach to occur with probability
\begin{equation}\label{eqn:p tide early}
  p_\textrm{tide}(a_\bin,y) = \frac{12a_0y r_*}{a_\bin^2},
\end{equation}
where $1 < y$ is a free parameter, which describes, roughly, the maximum separation between two bodies which would engender sizeable tidal effects. Equation \eqref{eqn:p tide} is strictly correct in the limit where $yr_* \ll a_\bin$ \citep{HutInagaki1985}, which is the limit we consider here; at larger $yr_*$ it has to be modified \citep{HutInagaki1985,SigurdssonPhinney1993}. It originates from the following reasoning: let $u < y$. For $ur_* \ll a_\bin$, the cross-section for star $1$ (which we choose to be the white dwarf) to come within a distance $ur_*$ of one of the other two stars is approximately described by a two-body interaction with either one of them. Gravitational focussing thus implies that the cross-section for this event is $4\pi Gmur_*/v^2$, where $v^2$ is the ``initial'' velocity of star $1$. As the tidal interaction occurs during the chaotic 3-body-close-interaction phase, $v$ should be, roughly, given by the virial speed, multiplied by $\sqrt{2}$ because it is a relative velocity, i.e. $v^2 \approx \frac{4}{3}\frac{\abs{E}}{m}$, with $E \approx - \frac{Gm^2}{2a_0}$, since the original binary was hard. The cross-section should be divided by the total area available for star $1$, which is approximately $\pi R^2$, and multiplied by two to account for the two possible partners $2$ and $3$. Thus,
\begin{equation}\label{eqn:p tide}
  p_\tide(a_\bin,u) = 2\times\frac{4\pi Gmur_*}{\pi R^2}\frac{3a_0}{2Gm} = \frac{12a_0u r_*}{R^2},
\end{equation}
as in equation \eqref{eqn:p tide early}. $p_\tide(a_\bin,u)$ is therefore the probability of star $1$ coming within $ur_*$ from star $2$ or star $3$. The probability density function is
\begin{equation}
  \frac{\mathrm{d}p_\tide}{\mathrm{d}u} = \begin{cases}
                                            \frac{12a_0r_*}{R^2}, & \mbox{if } 0 < u < \frac{R^2}{12a_0r_*} \\
                                            0, & \mbox{otherwise}.
                                          \end{cases}.
\end{equation}

If such an event does occur (i.e. if star $1$ comes to $(u+\mathrm{d}u)r_*$ from either of its companions, but not below $ur_*$), the energy loss is given by (see ref. \cite{PressTeukolsky1977}, with $T_2(x) \sim x^{8/3}$)
\begin{equation}\label{eqn:delta_E_tide_PT1977}
  \Delta E(u) = -2.995\frac{Gm^2}{r_*}u^{-10}.
\end{equation}
To gauge the error on our computations, we also use the simpler model of ref. \cite{Fabianetal1975}, in which
\begin{equation}\label{eqn:delta_E_tide_FPR1975}
  \Delta E(u) = -\frac{Gm^2}{r_*}u^{-6}.
\end{equation}
Needless to say, the technique described in \S \ref{sec:random walks} applies to more sophisticated tidal models, too.

\subsection{Perturbative Calculation of $P(x)$}
Let us define the following functions of the total triple energy $E$ (and, albeit suppressed, total angular momentum $\mathbf{J}$), for an initial total energy $E_i$ (please bear in mind that $\Delta E < 0$):
\begin{widetext}
\begin{align}
  & p(E,u) f_\bin(E_\bin,\mathbf{S}|E,\mathbf{J})\delta(E-(E_i+\Delta E(u))) = \int_{\set{E_\bin' \geq E'}} \mathrm{d}x'h(x|x')\frac{\mathrm{d}p_\tide(E_\bin')}{\mathrm{d}u} \\ &
  q(E)f_\bin(E_\bin,\mathbf{S}|E,\mathbf{J})\delta(E-E_i) = \int_{\set{E_\bin' \geq E'}} \mathrm{d}x'h(x|x')(1-p_\tide(E_b',y))
\end{align}

Suppose that $r_* \ll a_\bin$. Then, $p_\tide \ll 1$, and most of the steps conserve the total binary energy. Then one can view this process as two random walks: a ``small-scale'' random walk in $E_\bin$, and a larger-scale one in $E$, on top of it, and derive the equation perturbatively. What this means is that, in expanding $P_n$, one may truncate the series at some small power of $\eps =\frac{r_* \abs{E_i}}{GM\mu_s}$, where $E_i$ is the initial total energy. Such a truncation corresponds to \emph{re-summing} the series of $P(x)$, as a series in powers of $\eps$. Suppose one stops at first order; then
\begin{equation}
\begin{aligned}
  P_n(E_\bin,E|E_{b,i},E_i) & = \int_0^y \mathrm{d}u\bigg\{W_{\textrm{unlim}} \bigg[f(E_\bin|E)\bigg(\delta(E-E_i)\frac{q_\tide(a_0)}{y}q(E_i)^{n-1} \\
   & + ~\delta(E-E_i+\Delta E(u))\sum_{k=1}^{n-1}q_\tide(a_0)q(E_i)^{k-2}p(E_i,u)q(E_i - \Delta E(u))^{n-k} \\
  & + \left.\left.\left.\delta(E-E_i+\Delta E(u))\frac{\mathrm{d}p_\tide(a_0,u)}{\mathrm{d}u}q(E_i - \Delta E(u))^{n-1}\right)\right]\right\}
\end{aligned}
\end{equation}
The first line in this equation corresponds to the occurrence of no tidal interactions, the last -- to a tidal interaction in the very first close approach, and the second -- to a tidal interaction in another close approach.

The sum over $k$ is a geometric sum, and may be computed analytically. Then, using equation \eqref{eqn:random walk final probability} (summing from $n=1$, as we assume that the first, initial close approach always happens), one may sum over $n$ analytically, too (this sum may be exchanged with the action of $W_\textrm{unlim}$ by its linearity). Then one may act with $W_\textrm{unlim}$ to find that the probability distribution for the final binary energy reads
\begin{equation}\label{eqn: probability tides final}
\begin{aligned}
  P(E_b,\mathbf{S},E|E_i,\mathbf{J}) & \propto f_\bin(E_b,\mathbf{S}|E,\mathbf{J})\\ &
  \times \left\{\frac{q_\tide(a_0)\delta(E - E_i)}{1-q(E_i)} + \int_{0}^{y}\mathrm{d}u\frac{\delta(E - E_i + \Delta E(u))}{1-q(E_i - \Delta E(u))}\left[\frac{q_\textrm{tide}(a_0)p(E_i,u)}{1-q(E_i)} + \frac{\mathrm{d}p_\tide(a_0,u)}{\mathrm{d}u}\right]\right\}
\end{aligned}
\end{equation}
up to an overall normalisation and $O(\eps^2)$ corrections.
\end{widetext}
\begin{figure*}
  \centering
  \includegraphics[width=0.9\textwidth]{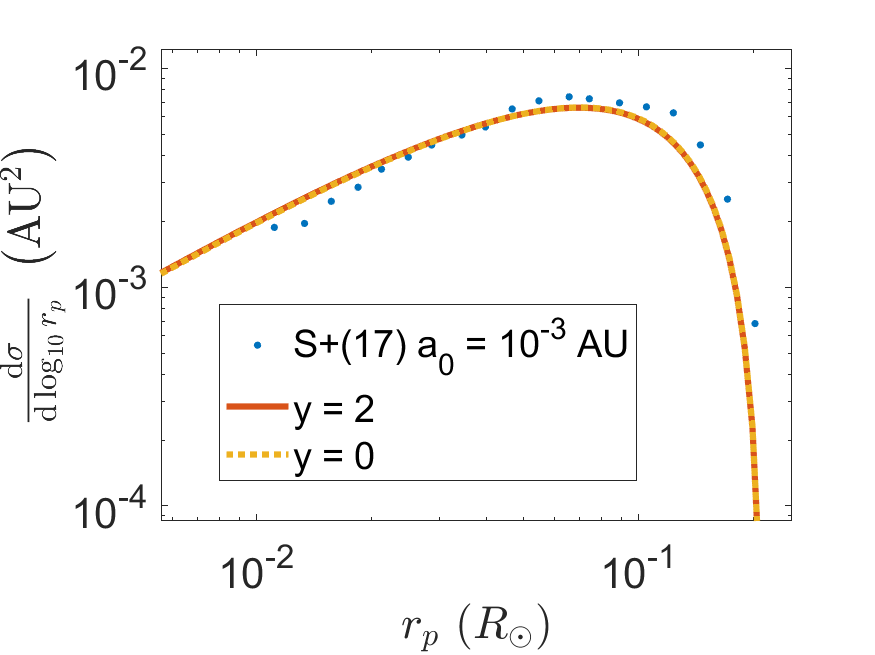}
  \caption{The differential cross-section for obtaining a final pericentre distance $r_p$ in an exchange/fly-by, which is the integral over $(E_\bin,\mathbf{S})$ of equation \eqref{eqn: probability tides final}, multiplied by the relevant Dirac delta-function. $y$ is defined below equation \eqref{eqn:p tide early}. This is compared with data from \citet{Samsingetal2017} for $a_0 = 10^{-3}~\textrm{AU}$, and $v_0 = 10 ~\textrm{km s}^{-1}$, and both are normalised to give the same total exchange/fly-by cross-section. To reduce computational difficulty here, $\Delta E(u)$ was approximated to be equal to $\Delta E((1+y)/2)$ for $u < y$, and zero otherwise. The $y = 2$ case is indiscernible from tide-less case, $y=0$, in agreement with \citeauthor{Samsingetal2017}'s result that tides do not change the exchange/fly-by cross-section significantly.}\label{fig:peri tides}
\end{figure*}

\subsection{In-Spiral Cross-Section}
\label{subsec:in-spiral cross-section}
\citet{Samsingetal2017} define the result of an encounter to be deemed `an in-spiral' if $a_\bin \leq 6r_*$ and if a collision has not occurred. The final possible outcomes are therefore: a collision, an in-spiral, an exchange or a fly-by. The exchange/fly-by cross-section may be computed by including a Heaviside function in $p,q,p_u,q_u$ which ensures that there is no collision and no in-spiral. The in-spiral/collision cross-section is simply the total cross-section, minus the exchange/fly-by cross-section.

There is another possible outcome: if $\Delta E(u)$ is large enough relative to $E_i$, then for large enough $J$, upon losing energy to tides, there is too much angular momentum for the triple system to interact closely again -- this manifests itself in none of the conditions in \S \ref{subsec: Celtica in angle-action variables} being satisfied. Denote the minimum such $J$ by $J_*$. In this case, the triple becomes hierarchical automatically, and the encounter ends. The inevitable fate of this triple is a tidal in-spiral: during each pericentre approach of the inner binary more energy is lost to tides, until the two stars collide. This is just another route to a tidal in-spiral, which does not require the triple to have ejected the third star.

\begin{figure}
  \centering
  \includegraphics[width=0.45\textwidth]{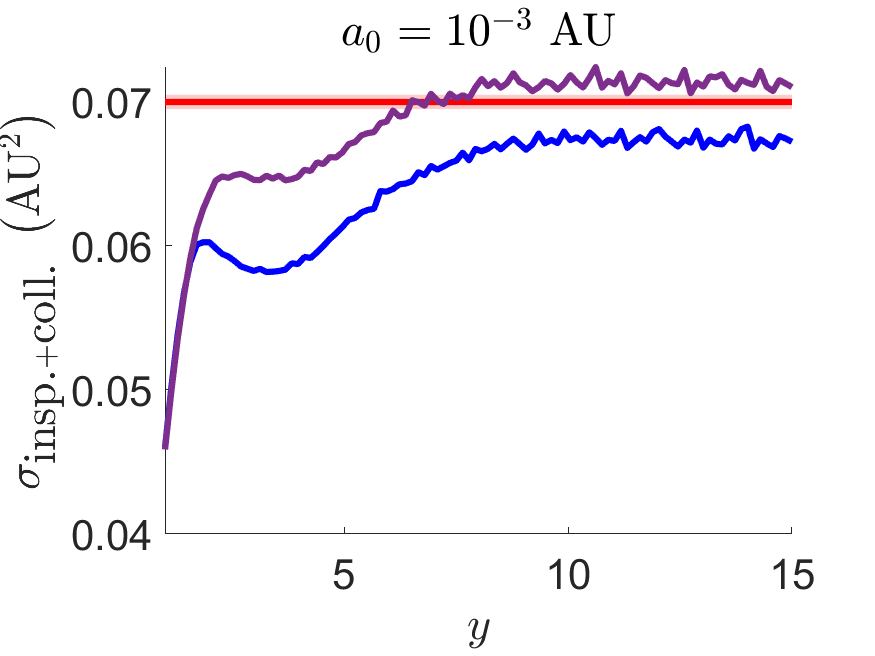}
  \caption{The in-spiral/collision cross-section, as a function of $y$, for $\beta = 1.3$, $a_0 = 10^{-3}$ AU and $v_0 = 10~\textrm{km s}^{-1}$. The purple line corresponds to the model of equation \eqref{eqn:delta_E_tide_FPR1975}, while the blue line uses equation \eqref{eqn:delta_E_tide_PT1977}; the red line is the numerical result of \citet{Samsingetal2017}. As expected, the cross-section converges to a constant for large-enough $y$, despite some noise due to numerical integration.}\label{fig:sigma of y}
\end{figure}

As in both models $\Delta E(u)$ decays quite fast with $u$, the dependence on $y$ is very weak, and the cross-sections converge for sufficiently large $y$ (see figure \ref{fig:sigma of y}).

One may compute the tidal in-spiral cross-section as follows: compute the cross-section for there being no collision and no tidal in-spiral, and then subtract the result from the total cross-section, which we take, in this section, to be
\begin{equation}
  \sigma_\textrm{tot} = \frac{\pi GMa_0}{v_0^2}
\end{equation}
-- and not twice this value -- to be consistent with ref. \cite{Samsingetal2017}. The former is given by first computing the differential cross-section for a final periapsis $r_p$, by integrating equation \eqref{eqn: probability tides final} over the disc $D(v_0)$ -- this is effectively an integration over the allowed range of total angular momenta -- as well as over the binary actions with a Dirac delta function $\delta(r_p - a_\bin(1-e_\bin))$, all the while enforcing the no-in-spiral-and-no-collisions condition both in equation \eqref{eqn: probability tides final} and in the definitions of $p(E),q(E),p_u(E),q_u(E)$, by restricting the integration there to the domain $\set{a_\bin' \geq 6r_*}$. This differential cross-section is shown in figure \ref{fig:peri tides}. In this figure only, it does not matter if one sets $R = \beta a_\bin$ for simplicity -- this deviation from equation \eqref{eqn: R definition} does not change the results significantly, since all three masses are equal, whence equation \eqref{eqn: R definition} reduces to $R =\beta\min\set{a_0^{1/3}a_\bin^{2/3},a_\bin}$, and the distribution of $a_\bin$ is peaked sharply about $a_0$. The total exchange/fly-by cross-section is then found by integrating this differential cross-section over $r_p \geq r_*$ (to preclude collisions); by the special nature of equation \eqref{eqn:encounters final probability}, this automatically precludes collisions during all close approaches, not just the final one.

Explicitly, let
\begin{widetext}
\begin{align}
  & N_\textrm{ubd}(E,\mathbf{J}) = \int \mathrm{d}E_\bin\mathrm{d}^2\mathbf{S}f_\bin(E_\bin,\mathbf{S}|E,\mathbf{J}), \\ &
  N_\textrm{ubd}^{ni}(E,\mathbf{J}) = \int \mathrm{d}E_\bin\mathrm{d}^2\mathbf{S}f_\bin(E_\bin,\mathbf{S}|E,\mathbf{J})\Theta\left(\frac{Gm_\bin\mu_\bin}{2\times 6r_*} - E_\bin \right);
\end{align}
these quantities, when evaluated at $E = E_i - \Delta E$, are only defined for $J \leq J_*$, so for $J > J_*$ we define them to be zero. Further define
\begin{equation}
  B = \int_{D(v_0)} \mathrm{d}^2\mathbf{b} \left[\frac{\mathrm{d}p_\tide(a_0)}{\mathrm{d}u} + \frac{q_\tide(a_0)p(E_i,u)}{1-q(E_i)}\right]\times\Theta\left(J(\mathbf{b}) - J_*\right),
\end{equation}
where $J(\mathbf{b})$ is the magnitude of the total angular momentum as a function of the position $\mathbf{b}$ of the in-coming star on the disc $D(v_0)$. It follows from equation \eqref{eqn: probability tides final} that if we denote
\begin{align}
  & U = \int_{D(v_0)} \mathrm{d}^2\mathbf{b}\frac{N_\textrm{ubd}^{ni}(E_i,\mathbf{J})q_\tide}{1-q_{ni}(E_i)} + \frac{N_\textrm{ubd}^{ni}(E_i-\Delta E,\mathbf{J})}{1-q_{ni}(E_i-\Delta E)}\left(p_{\tide,nc}(a_0) + \frac{q_\tide(a_0)p_{ni}(E_i)}{1-q_{ni}(E_i)}\right), \\ &
  V = \int_{D(v_0)} \mathrm{d}^2\mathbf{b}\frac{N_\textrm{ubd}(E_i,\mathbf{J})q_\tide}{1-q(E_i)} + \frac{N_\textrm{ubd}(E_i-\Delta E,\mathbf{J})}{1-q(E_i-\Delta E)}\left(p_{\tide}(a_0) + \frac{q_\tide(a_0)p(E_i)}{1-q(E_i)}\right),
\end{align}
\end{widetext}
where $p_{\tide,nc}(a_\bin) = 12a_0(y-1)r_*/a_\bin^2$, to preclude collisions; $p_{ni}(E)$ is the analogue of $p(E)$, but with the same Heaviside theta function inserted as in the definition of $N_\textrm{ubd}^{ni}$ and with $p_{\tide,nc}(a_\bin)$ used instead of $p_\tide$; and $q_{ni}(E)$ is the same as $q(E)$, but also with the afore-mentioned Heaviside function,---if so, then the in-spiral/collision cross-section is given by
\begin{equation}\label{eqn:cross-section for tidal in-spiral of collision}
  \sigma_{\textrm{insp}+\textrm{coll}} = \sigma_{\textrm{tot}}\left(1-\frac{U}{V+B}\right).
\end{equation}

\begin{figure}
  \centering
  \includegraphics[width=0.45\textwidth]{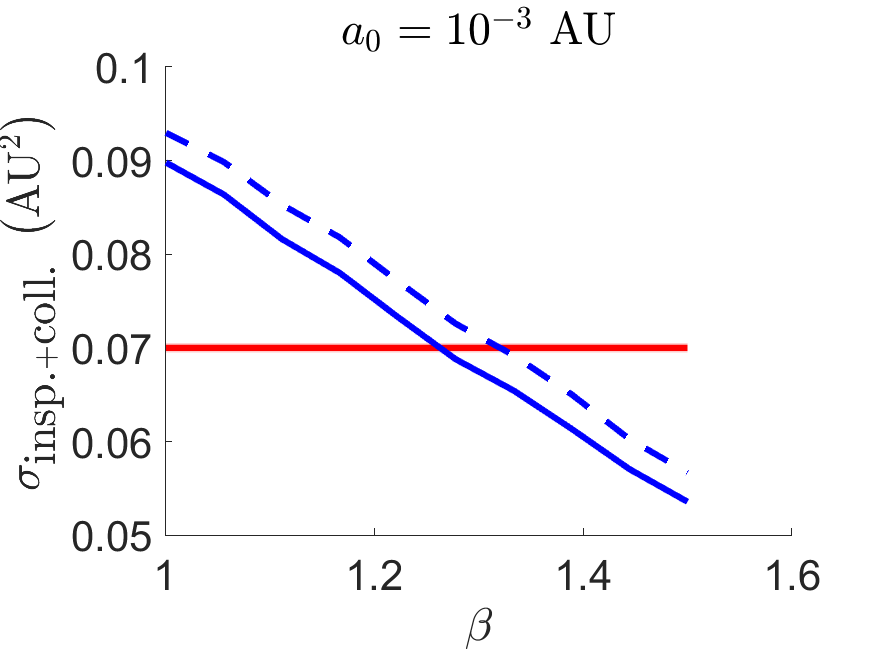}
  \includegraphics[width=0.45\textwidth]{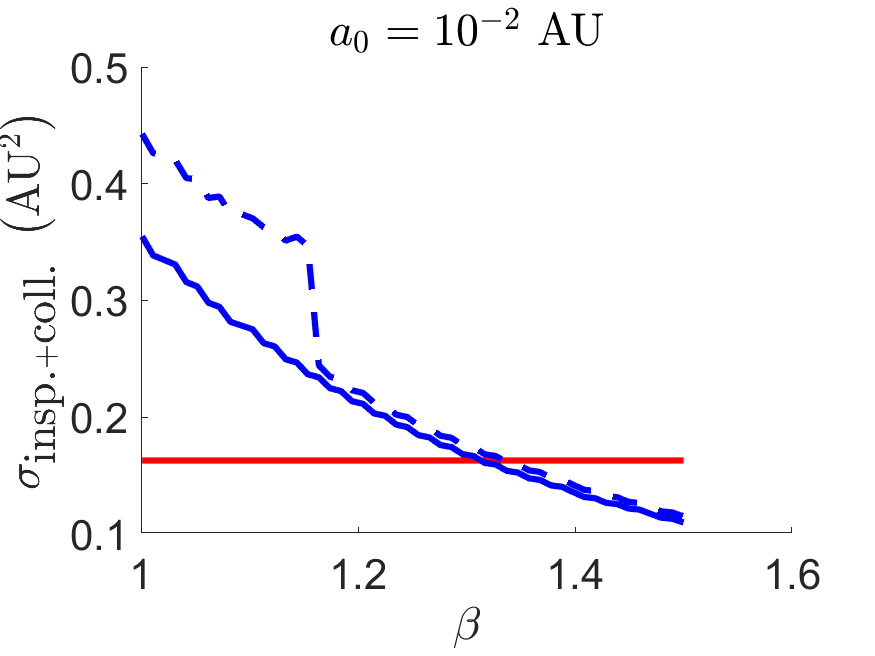}
  \caption{The in-spiral/collision cross-section for the case described in the text, for different values of $\beta$ (defined in equation \eqref{eqn: R definition}). The red line is the result of the simulations of \citet{Samsingetal2017}, the blue line shows our result, using the tidal model of equation \eqref{eqn:delta_E_tide_PT1977}, while the dashed blue line shows the cross-section if the model in equation \eqref{eqn:delta_E_tide_FPR1975} is employed. There is some noise due to numerical integration. Top: the case $a_0 = 10^{-3}$ AU; bottom: the case $a_0 = 10^{-2}$ AU. In both cases $v_0 = 10~\textrm{km s}^{-1}$, and $y$ is chosen to be very large.}\label{fig:sigma_insp of beta}
\end{figure}

We find, from equation \eqref{eqn:cross-section for tidal in-spiral of collision}, that for $\beta \approx 1.3$, $\sigma_{\textrm{insp}+\textrm{coll}}$, agrees with the findings of ref. \cite{Samsingetal2017} for both $a_0 = 10^{-3}$ and $10^{-2}$ AU. The former improves upon their analytic estimate of $0.155~\textrm{AU}^2$ considerably. This cross-section depends on $\beta$, as shown in figure \ref{fig:sigma_insp of beta}. The fact that $\beta = 1.3$ fits both cases, with initial semi-major axes differing by an order of magnitude, implies that indeed equation \eqref{eqn:cross-section for tidal in-spiral of collision} is an adequate model to describe a binary-single encounter with tides.

\section{Discussion and Summary}

In this paper we introduced a random walk model for binary-single encounters in globular clusters. An encounter is viewed as a sequence of chaotic, close triple approaches, interspersed with hierarchical phases. The orbital parameters of the binary and the single star, as well as the triple system's constants of motion, perform a random walk: each step of the walk corresponds to one close approach and its subsequent hierarchical phase. We calculated the transition probabilities between steps of the walk, both in the Newtonian, point-mass approximation (in which the walk is memory-less, and the final outcome distribution is simply given by the formula of ref. \cite{StoneLeigh2019}, while the transition probabilities between intermediate steps are given by equation \eqref{eqn:f_bin}), and in the case where there is some dissipative process involved, when equation \eqref{eqn:encounters final probability} holds.

We have shown that this model reproduces numerical results well, as we exemplified for aspects such as the semi-major axis distribution, the escaper's mass distribution, and the final periapsis distribution. Including tides and collisions, our predictions match the in-spiral/collision cross-section measured by numerical simulations. This validates the prescription of $R$ and the solution to the bound problem; besides, including a tides allowed us to perform a non-trivial test on the extra step involved in elevating the statistical solution of the scattering problem to the random walk model, which it passed.

In some cases the probability distribution can be computed completely analytically, while in others it only involves a relatively simple calculation of a few integrals. Please note, that the formula for the escaper's mass distribution is valid in every intermediate step of the encounter. In conjunction with the random walk model, it implies, \emph{inter alia}, that if one of the stars is considerably lighter than the other two, then it will be the one to be ejected after each close approach -- not just the final one. The only free parameters in our analysis, which do not influence the results of the dissipation-free problem significantly, are $\beta$ and $\eta$; the former of which was found to be equal to $1.3$ -- a single value that agrees with both cases considered in \S \ref{sec:numerical simulations}, and the latter does not change the results much in either case. The fact that one value of $\beta$ fits both supports the random walk approach further.

The random walk model described here may be used to address a plethora of astrophysical phenomena using the analytical, statistical model we described in this paper: apart from incorporating tidal interactions as was done here, one could also include gravitational-wave dissipation, and the effects of stellar evolution. One could also investigate external effects, for example the tidal influence of an external gravitational potential. The main change would be a modification of the largest value $a_s(1+e_s)$ that corresponds to a bound binary. Such cut-offs are important in globular clusters, which are the main places where one expects to find significant rates of binary-single encounters. This is made possible by our statistical approximate solution of the bound, non-hierarchical, three-body problem in equation \eqref{eqn:f_bin}.
This approach also allows one to calculate the distribution of the number of consecutive close approaches before the disruption of the triple, and hence the distribution of time-scales of temporary captures which could be relevant for various astrophysical capture processes by gravitating stars and planets. Other applications include -- but are not limited to -- binary-single encounters in nuclear star clusters around super-massive black holes at the centres of galaxies (or, equivalently, binary-single encounters of planets, dwarf planets, moons or asteroids in the solar system), where the Hill radius effectively provides a limiting separation (as mentioned above) during consecutive encounters, as well as the velocity distribution of fast, runaway stars due to ejections through binary-single encounters, which will be treated in a separate paper.

Let us present an algorithm for how this would be done for \emph{any} given astrophysical process involved in three-body physics. Suppose, that in addition to Newtonian, point-particle motion one wishes to incorporate another physical phenomenon, say, gravitational-wave emission (which we will address in future work). One would have to be able to calculate how this phenomenon changes the total energy and angular momentum $E$, $\mathbf{J}$ during a single close approach and subsequent hierarchical phase. With these data, one could compute $f_c(E,\mathbf{J}|E',E_\bin',\mathbf{S}',\mathbf{J}')$ from \S \ref{sec:random walks}. Using the random walk model, it gives the transition probabilities $h(x|x')$ immediately, and then all that remains is to sum up the series \eqref{eqn:random walk final probability} to obtain the final probability -- the probability density function of final binary parameters, given initial total energy and angular momentum. If there is a small parameter, e.g. if the probability for non-zero change in the constants of motion is small, one can re-sum equation \eqref{eqn:random walk final probability}, and expand in the small parameter. To summarise:
\begin{enumerate}
  \item Compute the single-step (one close approach and one hierarchical phase) probabilities for a given change in total energy and angular momentum, $f_c(E,\mathbf{J}|E',E_\bin',\mathbf{S}',\mathbf{J}')$;
  \item compute the transition probabilities $h(x|x')$ as in \S \ref{sec:random walks};
  \item and compute equation \eqref{eqn:random walk final probability} to obtain the final distribution.
  \item If $h(x|x')$ is small ($O(\eps)$) for $(E',\mathbf{J}') \neq (E,\mathbf{J})$, re-sum equation \eqref{eqn:random walk final probability} and expand in $\eps$ to the desired accuracy.
\end{enumerate}

In our paper we brought the endeavour of statistical modelling of binary-single encounters closer to completion. By viewing them as concatenations of close triple approaches, we were able to model, statistically, the \emph{bound} non-hierarchical three-body problem, and to use the solution to model the entire encounter as a random walk. This model has the potential to facilitate simulations of globular clusters significantly. Instead of having to resolve binary-single encounters individually using a high-resolution few-body code, one could simply implement the analytical random-walk model as a probabilistic solution of these encounters, with the additional possibility of incorporating any astrophysical process one wants. By the law of large numbers, if the number of encounters per cluster is sufficiently large, the few-body resolutions of binary-single encounters will be rendered unnecessary. We hope that this gain in speed would enable astrophysicists to study many more phenomena in clusters, planetary systems and in the field in much better detail.

\section*{Acknowledgments}

We wish to thank Melvyn Davis, Vincent Desjacques, Evgeni Grishin, Barak Kol, Nathan Leigh, Ilya Mandel, Mor Rozner, Johan Samsing, Steinn Sigurdsson and Nicholas Stone for helpful discussions. HBP would like to thank the kind support from the Kingsley distinguished-visitor program in Caltech, where some of the work was done. HBP and YBG acknowledge support for this project from the European Union's Horizon 2020 research and innovation program under grant agreement No 865932-ERC-SNeX. YBG also acknowledges support by the Israel Science Foundation (grant no. 1395/16) and by the Israeli Academy of Sciences' Adams Fellowship.

\bibliography{encounters,bibliography}

\begin{thebibliography}{37}%
\makeatletter
\providecommand \@ifxundefined [1]{%
 \@ifx{#1\undefined}
}%
\providecommand \@ifnum [1]{%
 \ifnum #1\expandafter \@firstoftwo
 \else \expandafter \@secondoftwo
 \fi
}%
\providecommand \@ifx [1]{%
 \ifx #1\expandafter \@firstoftwo
 \else \expandafter \@secondoftwo
 \fi
}%
\providecommand \natexlab [1]{#1}%
\providecommand \enquote  [1]{``#1''}%
\providecommand \bibnamefont  [1]{#1}%
\providecommand \bibfnamefont [1]{#1}%
\providecommand \citenamefont [1]{#1}%
\providecommand \href@noop [0]{\@secondoftwo}%
\providecommand \href [0]{\begingroup \@sanitize@url \@href}%
\providecommand \@href[1]{\@@startlink{#1}\@@href}%
\providecommand \@@href[1]{\endgroup#1\@@endlink}%
\providecommand \@sanitize@url [0]{\catcode `\\12\catcode `\$12\catcode
  `\&12\catcode `\#12\catcode `\^12\catcode `\_12\catcode `\%12\relax}%
\providecommand \@@startlink[1]{}%
\providecommand \@@endlink[0]{}%
\providecommand \url  [0]{\begingroup\@sanitize@url \@url }%
\providecommand \@url [1]{\endgroup\@href {#1}{\urlprefix }}%
\providecommand \urlprefix  [0]{URL }%
\providecommand \Eprint [0]{\href }%
\providecommand \doibase [0]{https://doi.org/}%
\providecommand \selectlanguage [0]{\@gobble}%
\providecommand \bibinfo  [0]{\@secondoftwo}%
\providecommand \bibfield  [0]{\@secondoftwo}%
\providecommand \translation [1]{[#1]}%
\providecommand \BibitemOpen [0]{}%
\providecommand \bibitemStop [0]{}%
\providecommand \bibitemNoStop [0]{.\EOS\space}%
\providecommand \EOS [0]{\spacefactor3000\relax}%
\providecommand \BibitemShut  [1]{\csname bibitem#1\endcsname}%
\let\auto@bib@innerbib\@empty
\bibitem [{\citenamefont {{Binney}}\ and\ \citenamefont
  {{Tremaine}}(2008)}]{Binney}%
  \BibitemOpen
  \bibfield  {author} {\bibinfo {author} {\bibfnamefont {J.}~\bibnamefont
  {{Binney}}}\ and\ \bibinfo {author} {\bibfnamefont {S.}~\bibnamefont
  {{Tremaine}}},\ }\href@noop {} {\emph {\bibinfo {title} {Galactic Dynamics:
  Second Edition, by James Binney and Scott Tremaine.~ISBN 978-0-691-13026-2
  (HB).~Published by Princeton University Press, Princeton, NJ USA, 2008.}}}\
  (\bibinfo  {publisher} {Princeton University Press},\ \bibinfo {year}
  {2008})\BibitemShut {NoStop}%
\bibitem [{\citenamefont {Heggie}\ and\ \citenamefont
  {Hut}(2003)}]{HeggieHut2003}%
  \BibitemOpen
  \bibfield  {author} {\bibinfo {author} {\bibfnamefont {D.}~\bibnamefont
  {Heggie}}\ and\ \bibinfo {author} {\bibfnamefont {P.}~\bibnamefont {Hut}},\
  }\href {https://doi.org/10.1017/CBO9781139164535} {\emph {\bibinfo {title}
  {The Gravitational Million–Body Problem: A Multidisciplinary Approach to
  Star Cluster Dynamics}}}\ (\bibinfo  {publisher} {Cambridge University Press,
  Cambridge},\ \bibinfo {year} {2003})\BibitemShut {NoStop}%
\bibitem [{\citenamefont {{Perets}}\ and\ \citenamefont
  {{Kratter}}(2012)}]{Per+12}%
  \BibitemOpen
  \bibfield  {author} {\bibinfo {author} {\bibfnamefont {H.~B.}\ \bibnamefont
  {{Perets}}}\ and\ \bibinfo {author} {\bibfnamefont {K.~M.}\ \bibnamefont
  {{Kratter}}},\ }\bibfield  {title} {\bibinfo {title} {{The Triple Evolution
  Dynamical Instability: Stellar Collisions in the Field and the Formation of
  Exotic Binaries}},\ }\href {https://doi.org/10.1088/0004-637X/760/2/99}
  {\bibfield  {journal} {\bibinfo  {journal} {\apj}\ }\textbf {\bibinfo
  {volume} {760}},\ \bibinfo {eid} {99} (\bibinfo {year} {2012})},\ \Eprint
  {https://arxiv.org/abs/1203.2914} {arXiv:1203.2914 [astro-ph.SR]}
  \BibitemShut {NoStop}%
\bibitem [{\citenamefont {{Michaely}}\ and\ \citenamefont
  {{Perets}}(2020)}]{Mic+20}%
  \BibitemOpen
  \bibfield  {author} {\bibinfo {author} {\bibfnamefont {E.}~\bibnamefont
  {{Michaely}}}\ and\ \bibinfo {author} {\bibfnamefont {H.~B.}\ \bibnamefont
  {{Perets}}},\ }\bibfield  {title} {\bibinfo {title} {{High rate of
  gravitational waves mergers from flyby perturbations of wide black-hole
  triples in the field}},\ }\bibfield  {journal} {\bibinfo  {journal} {\mnras}\
  }\href {https://doi.org/10.1093/mnras/staa2720} {10.1093/mnras/staa2720}
  (\bibinfo {year} {2020}),\ \Eprint {https://arxiv.org/abs/2008.01094}
  {arXiv:2008.01094 [astro-ph.HE]} \BibitemShut {NoStop}%
\bibitem [{\citenamefont {{Heggie}}(1975)}]{Heggie1975}%
  \BibitemOpen
  \bibfield  {author} {\bibinfo {author} {\bibfnamefont {D.~C.}\ \bibnamefont
  {{Heggie}}},\ }\bibfield  {title} {\bibinfo {title} {{Binary evolution in
  stellar dynamics.}},\ }\href {https://doi.org/10.1093/mnras/173.3.729}
  {\bibfield  {journal} {\bibinfo  {journal} {\mnras}\ }\textbf {\bibinfo
  {volume} {173}},\ \bibinfo {pages} {729} (\bibinfo {year}
  {1975})}\BibitemShut {NoStop}%
\bibitem [{\citenamefont {{Valtonen}}\ and\ \citenamefont
  {{Karttunen}}(2006)}]{ValtonenKarttunen2006}%
  \BibitemOpen
  \bibfield  {author} {\bibinfo {author} {\bibfnamefont {M.}~\bibnamefont
  {{Valtonen}}}\ and\ \bibinfo {author} {\bibfnamefont {H.}~\bibnamefont
  {{Karttunen}}},\ }\href@noop {} {\emph {\bibinfo {title} {{The Three-Body
  Problem}}}}\ (\bibinfo  {publisher} {Cambridge University Press, Cambridge},\
  \bibinfo {year} {2006})\BibitemShut {NoStop}%
\bibitem [{\citenamefont {Arnold}\ \emph {et~al.}(2006)\citenamefont {Arnold},
  \citenamefont {Kozlov},\ and\ \citenamefont {Neishtadt}}]{Arnoldetal2006}%
  \BibitemOpen
  \bibfield  {author} {\bibinfo {author} {\bibfnamefont {V.~I.}\ \bibnamefont
  {Arnold}}, \bibinfo {author} {\bibfnamefont {V.~V.}\ \bibnamefont {Kozlov}},\
  and\ \bibinfo {author} {\bibfnamefont {A.~I.}\ \bibnamefont {Neishtadt}},\
  }\href@noop {} {\emph {\bibinfo {title} {Mathematical aspects of classical
  and celestial mechanics}}},\ \bibinfo {edition} {3rd}\ ed.,\ \bibinfo
  {series} {Encyclopaedia of Mathematical Sciences}, Vol.~\bibinfo {volume}
  {3}\ (\bibinfo  {publisher} {Springer-Verlag, Berlin},\ \bibinfo {year}
  {2006})\ pp.\ \bibinfo {pages} {xiv+518},\ \bibinfo {note} {[Dynamical
  systems. III], Translated from the Russian original by E.
  Khukhro}\BibitemShut {NoStop}%
\bibitem [{\citenamefont {{Saslaw}}\ \emph {et~al.}(1974)\citenamefont
  {{Saslaw}}, \citenamefont {{Valtonen}},\ and\ \citenamefont
  {{Aarseth}}}]{Saslawetal1974}%
  \BibitemOpen
  \bibfield  {author} {\bibinfo {author} {\bibfnamefont {W.~C.}\ \bibnamefont
  {{Saslaw}}}, \bibinfo {author} {\bibfnamefont {M.~J.}\ \bibnamefont
  {{Valtonen}}},\ and\ \bibinfo {author} {\bibfnamefont {S.~J.}\ \bibnamefont
  {{Aarseth}}},\ }\bibfield  {title} {\bibinfo {title} {{The Gravitational
  Slingshot and the Structure of Extragalactic Radio Sources}},\ }\href
  {https://doi.org/10.1086/152870} {\bibfield  {journal} {\bibinfo  {journal}
  {\apj}\ }\textbf {\bibinfo {volume} {190}},\ \bibinfo {pages} {253} (\bibinfo
  {year} {1974})}\BibitemShut {NoStop}%
\bibitem [{\citenamefont {{Hills}}(1975)}]{Hills1975}%
  \BibitemOpen
  \bibfield  {author} {\bibinfo {author} {\bibfnamefont {J.~G.}\ \bibnamefont
  {{Hills}}},\ }\bibfield  {title} {\bibinfo {title} {{Encounters between
  binary and single stars and their effect on the dynamical evolution of
  stellar systems.}},\ }\href {https://doi.org/10.1086/111815} {\bibfield
  {journal} {\bibinfo  {journal} {\aj}\ }\textbf {\bibinfo {volume} {80}},\
  \bibinfo {pages} {809} (\bibinfo {year} {1975})}\BibitemShut {NoStop}%
\bibitem [{\citenamefont {{Hills}}\ and\ \citenamefont
  {{Fullerton}}(1980)}]{HillsFullerton1980}%
  \BibitemOpen
  \bibfield  {author} {\bibinfo {author} {\bibfnamefont {J.~G.}\ \bibnamefont
  {{Hills}}}\ and\ \bibinfo {author} {\bibfnamefont {L.~W.}\ \bibnamefont
  {{Fullerton}}},\ }\bibfield  {title} {\bibinfo {title} {{Computer simulations
  of close encounters between single stars and hard binaries}},\ }\href
  {https://doi.org/10.1086/112798} {\bibfield  {journal} {\bibinfo  {journal}
  {\aj}\ }\textbf {\bibinfo {volume} {85}},\ \bibinfo {pages} {1281} (\bibinfo
  {year} {1980})}\BibitemShut {NoStop}%
\bibitem [{\citenamefont {{Anosova}}(1986)}]{Anosova1986}%
  \BibitemOpen
  \bibfield  {author} {\bibinfo {author} {\bibfnamefont {J.~P.}\ \bibnamefont
  {{Anosova}}},\ }\bibfield  {title} {\bibinfo {title} {{Dynamical Evolution of
  Triple Systems}},\ }\href {https://doi.org/10.1007/BF00656037} {\bibfield
  {journal} {\bibinfo  {journal} {\apss}\ }\textbf {\bibinfo {volume} {124}},\
  \bibinfo {pages} {217} (\bibinfo {year} {1986})}\BibitemShut {NoStop}%
\bibitem [{\citenamefont {{Anosova}}\ and\ \citenamefont
  {{Orlov}}(1986)}]{AnosovaOrlov1986}%
  \BibitemOpen
  \bibfield  {author} {\bibinfo {author} {\bibfnamefont {Z.~P.}\ \bibnamefont
  {{Anosova}}}\ and\ \bibinfo {author} {\bibfnamefont {V.~V.}\ \bibnamefont
  {{Orlov}}},\ }\bibfield  {title} {\bibinfo {title} {{Dynamical Evolution of
  Equal-Mass Triple Systems in Three Dimensions}},\ }\href@noop {} {\bibfield
  {journal} {\bibinfo  {journal} {\sovast}\ }\textbf {\bibinfo {volume} {30}},\
  \bibinfo {pages} {380} (\bibinfo {year} {1986})}\BibitemShut {NoStop}%
\bibitem [{\citenamefont {{Hills}}(1989)}]{Hills1989}%
  \BibitemOpen
  \bibfield  {author} {\bibinfo {author} {\bibfnamefont {J.~G.}\ \bibnamefont
  {{Hills}}},\ }\bibfield  {title} {\bibinfo {title} {{Effect of Intruder Mass
  on Collisions with Hard Binaries. I. Zero-Impact Parameter}},\ }\href
  {https://doi.org/10.1086/114973} {\bibfield  {journal} {\bibinfo  {journal}
  {\aj}\ }\textbf {\bibinfo {volume} {97}},\ \bibinfo {pages} {222} (\bibinfo
  {year} {1989})}\BibitemShut {NoStop}%
\bibitem [{\citenamefont {{Hills}}(1992)}]{Hills1992}%
  \BibitemOpen
  \bibfield  {author} {\bibinfo {author} {\bibfnamefont {J.~G.}\ \bibnamefont
  {{Hills}}},\ }\bibfield  {title} {\bibinfo {title} {{Effects of Intruder Mass
  on Collisions With Hard Binaries. II. Dependence on Impact Parameter and
  Computations of the Interaction Cross Section}},\ }\href
  {https://doi.org/10.1086/116204} {\bibfield  {journal} {\bibinfo  {journal}
  {\aj}\ }\textbf {\bibinfo {volume} {103}},\ \bibinfo {pages} {1955} (\bibinfo
  {year} {1992})}\BibitemShut {NoStop}%
\bibitem [{\citenamefont {{Heggie}}\ and\ \citenamefont
  {{Hut}}(1993)}]{HeggieHut1993}%
  \BibitemOpen
  \bibfield  {author} {\bibinfo {author} {\bibfnamefont {D.~C.}\ \bibnamefont
  {{Heggie}}}\ and\ \bibinfo {author} {\bibfnamefont {P.}~\bibnamefont
  {{Hut}}},\ }\bibfield  {title} {\bibinfo {title} {{Binary--Single-Star
  Scattering. IV. Analytic Approximations and Fitting Formulae for Cross
  Sections and Reaction Rates}},\ }\href {https://doi.org/10.1086/191768}
  {\bibfield  {journal} {\bibinfo  {journal} {\apjs}\ }\textbf {\bibinfo
  {volume} {85}},\ \bibinfo {pages} {347} (\bibinfo {year} {1993})}\BibitemShut
  {NoStop}%
\bibitem [{\citenamefont {{Hut}}(1993)}]{Hut1993}%
  \BibitemOpen
  \bibfield  {author} {\bibinfo {author} {\bibfnamefont {P.}~\bibnamefont
  {{Hut}}},\ }\bibfield  {title} {\bibinfo {title} {{Binary--Single-Star
  Scattering. III. Numerical Experiments for Equal-Mass Hard Binaries}},\
  }\href {https://doi.org/10.1086/172199} {\bibfield  {journal} {\bibinfo
  {journal} {\apj}\ }\textbf {\bibinfo {volume} {403}},\ \bibinfo {pages} {256}
  (\bibinfo {year} {1993})}\BibitemShut {NoStop}%
\bibitem [{\citenamefont {{Sigurdsson}}\ and\ \citenamefont
  {{Phinney}}(1993)}]{SigurdssonPhinney1993}%
  \BibitemOpen
  \bibfield  {author} {\bibinfo {author} {\bibfnamefont {S.}~\bibnamefont
  {{Sigurdsson}}}\ and\ \bibinfo {author} {\bibfnamefont {E.~S.}\ \bibnamefont
  {{Phinney}}},\ }\bibfield  {title} {\bibinfo {title} {{Binary--Single Star
  Interactions in Globular Clusters}},\ }\href {https://doi.org/10.1086/173190}
  {\bibfield  {journal} {\bibinfo  {journal} {\apj}\ }\textbf {\bibinfo
  {volume} {415}},\ \bibinfo {pages} {631} (\bibinfo {year}
  {1993})}\BibitemShut {NoStop}%
\bibitem [{\citenamefont {{Mikkola}}(1994)}]{Mikkola1994}%
  \BibitemOpen
  \bibfield  {author} {\bibinfo {author} {\bibfnamefont {S.}~\bibnamefont
  {{Mikkola}}},\ }\bibfield  {title} {\bibinfo {title} {{A Numerical
  Exploration of the Phase-Space Structure of Chaotic Three-Body Scattering}},\
  }\href {https://doi.org/10.1093/mnras/269.1.127} {\bibfield  {journal}
  {\bibinfo  {journal} {\mnras}\ }\textbf {\bibinfo {volume} {269}},\ \bibinfo
  {pages} {127} (\bibinfo {year} {1994})}\BibitemShut {NoStop}%
\bibitem [{\citenamefont {{Samsing}}\ \emph {et~al.}(2014)\citenamefont
  {{Samsing}}, \citenamefont {{MacLeod}},\ and\ \citenamefont
  {{Ramirez-Ruiz}}}]{Samsingetal2014}%
  \BibitemOpen
  \bibfield  {author} {\bibinfo {author} {\bibfnamefont {J.}~\bibnamefont
  {{Samsing}}}, \bibinfo {author} {\bibfnamefont {M.}~\bibnamefont
  {{MacLeod}}},\ and\ \bibinfo {author} {\bibfnamefont {E.}~\bibnamefont
  {{Ramirez-Ruiz}}},\ }\bibfield  {title} {\bibinfo {title} {{The Formation of
  Eccentric Compact Binary Inspirals and the Role of Gravitational Wave
  Emission in Binary-Single Stellar Encounters}},\ }\href
  {https://doi.org/10.1088/0004-637X/784/1/71} {\bibfield  {journal} {\bibinfo
  {journal} {\apj}\ }\textbf {\bibinfo {volume} {784}},\ \bibinfo {eid} {71}
  (\bibinfo {year} {2014})},\ \Eprint {https://arxiv.org/abs/1308.2964}
  {arXiv:1308.2964 [astro-ph.HE]} \BibitemShut {NoStop}%
\bibitem [{\citenamefont {{Leigh}}\ and\ \citenamefont
  {{Wegsman}}(2018)}]{LeighWegsman2018}%
  \BibitemOpen
  \bibfield  {author} {\bibinfo {author} {\bibfnamefont {N.~W.~C.}\
  \bibnamefont {{Leigh}}}\ and\ \bibinfo {author} {\bibfnamefont
  {S.}~\bibnamefont {{Wegsman}}},\ }\bibfield  {title} {\bibinfo {title}
  {{Illustrating chaos: a schematic discretization of the general three-body
  problem in Newtonian gravity}},\ }\href
  {https://doi.org/10.1093/mnras/sty192} {\bibfield  {journal} {\bibinfo
  {journal} {\mnras}\ }\textbf {\bibinfo {volume} {476}},\ \bibinfo {pages}
  {336} (\bibinfo {year} {2018})},\ \Eprint {https://arxiv.org/abs/1801.07257}
  {arXiv:1801.07257 [astro-ph.SR]} \BibitemShut {NoStop}%
\bibitem [{\citenamefont {{Manwadkar}}\ \emph {et~al.}(2020)\citenamefont
  {{Manwadkar}}, \citenamefont {{Trani}},\ and\ \citenamefont
  {{Leigh}}}]{Manwadkaretal2020}%
  \BibitemOpen
  \bibfield  {author} {\bibinfo {author} {\bibfnamefont {V.}~\bibnamefont
  {{Manwadkar}}}, \bibinfo {author} {\bibfnamefont {A.~A.}\ \bibnamefont
  {{Trani}}},\ and\ \bibinfo {author} {\bibfnamefont {N.~W.~C.}\ \bibnamefont
  {{Leigh}}},\ }\bibfield  {title} {\bibinfo {title} {{Chaos and L{\'e}vy
  flights in the three-body problem}},\ }\href
  {https://doi.org/10.1093/mnras/staa1722} {\bibfield  {journal} {\bibinfo
  {journal} {\mnras}\ }\textbf {\bibinfo {volume} {497}},\ \bibinfo {pages}
  {3694} (\bibinfo {year} {2020})},\ \Eprint {https://arxiv.org/abs/2004.05475}
  {arXiv:2004.05475 [astro-ph.EP]} \BibitemShut {NoStop}%
\bibitem [{\citenamefont {{Monaghan}}(1976{\natexlab{a}})}]{Monaghan1976a}%
  \BibitemOpen
  \bibfield  {author} {\bibinfo {author} {\bibfnamefont {J.~J.}\ \bibnamefont
  {{Monaghan}}},\ }\bibfield  {title} {\bibinfo {title} {{A statistical theory
  of the disruption of three-body systems - I. Low angular momentum.}},\ }\href
  {https://doi.org/10.1093/mnras/176.1.63} {\bibfield  {journal} {\bibinfo
  {journal} {\mnras}\ }\textbf {\bibinfo {volume} {176}},\ \bibinfo {pages}
  {63} (\bibinfo {year} {1976}{\natexlab{a}})}\BibitemShut {NoStop}%
\bibitem [{\citenamefont {{Monaghan}}(1976{\natexlab{b}})}]{Monaghan1976b}%
  \BibitemOpen
  \bibfield  {author} {\bibinfo {author} {\bibfnamefont {J.~J.}\ \bibnamefont
  {{Monaghan}}},\ }\bibfield  {title} {\bibinfo {title} {{A statistical theory
  of the disruption of three-body systems - II. High angular momentum.}},\
  }\href {https://doi.org/10.1093/mnras/177.3.583} {\bibfield  {journal}
  {\bibinfo  {journal} {\mnras}\ }\textbf {\bibinfo {volume} {177}},\ \bibinfo
  {pages} {583} (\bibinfo {year} {1976}{\natexlab{b}})}\BibitemShut {NoStop}%
\bibitem [{\citenamefont {{Nash}}\ and\ \citenamefont
  {{Monaghan}}(1978)}]{NashMonaghan1978}%
  \BibitemOpen
  \bibfield  {author} {\bibinfo {author} {\bibfnamefont {P.~E.}\ \bibnamefont
  {{Nash}}}\ and\ \bibinfo {author} {\bibfnamefont {J.~J.}\ \bibnamefont
  {{Monaghan}}},\ }\bibfield  {title} {\bibinfo {title} {{A statistical theory
  of the disruption of three-body systems - III. Three-dimensional motion.}},\
  }\href {https://doi.org/10.1093/mnras/184.1.119} {\bibfield  {journal}
  {\bibinfo  {journal} {\mnras}\ }\textbf {\bibinfo {volume} {184}},\ \bibinfo
  {pages} {119} (\bibinfo {year} {1978})}\BibitemShut {NoStop}%
\bibitem [{\citenamefont {{Kol}}(2020)}]{Kol2020}%
  \BibitemOpen
  \bibfield  {author} {\bibinfo {author} {\bibfnamefont {B.}~\bibnamefont
  {{Kol}}},\ }\bibfield  {title} {\bibinfo {title} {{Flux-based statistical
  prediction of three-body outcomes}},\ }\href@noop {} {\bibfield  {journal}
  {\bibinfo  {journal} {arXiv e-prints}\ ,\ \bibinfo {eid} {arXiv:2002.11496}}
  (\bibinfo {year} {2020})},\ \Eprint {https://arxiv.org/abs/2002.11496}
  {arXiv:2002.11496 [gr-qc]} \BibitemShut {NoStop}%
\bibitem [{\citenamefont {{Stone}}\ and\ \citenamefont
  {{Leigh}}(2019)}]{StoneLeigh2019}%
  \BibitemOpen
  \bibfield  {author} {\bibinfo {author} {\bibfnamefont {N.~C.}\ \bibnamefont
  {{Stone}}}\ and\ \bibinfo {author} {\bibfnamefont {N.~W.~C.}\ \bibnamefont
  {{Leigh}}},\ }\bibfield  {title} {\bibinfo {title} {{A statistical solution
  to the chaotic, non-hierarchical three-body problem}},\ }\href
  {https://doi.org/10.1038/s41586-019-1833-8} {\bibfield  {journal} {\bibinfo
  {journal} {\nat}\ }\textbf {\bibinfo {volume} {576}},\ \bibinfo {pages} {406}
  (\bibinfo {year} {2019})}\BibitemShut {NoStop}%
\bibitem [{\citenamefont {{Hut}}\ and\ \citenamefont
  {{Inagaki}}(1985)}]{HutInagaki1985}%
  \BibitemOpen
  \bibfield  {author} {\bibinfo {author} {\bibfnamefont {P.}~\bibnamefont
  {{Hut}}}\ and\ \bibinfo {author} {\bibfnamefont {S.}~\bibnamefont
  {{Inagaki}}},\ }\bibfield  {title} {\bibinfo {title} {{Globular cluster
  evolution with finite-size stars - Cross sections and reaction rates}},\
  }\href {https://doi.org/10.1086/163636} {\bibfield  {journal} {\bibinfo
  {journal} {\apj}\ }\textbf {\bibinfo {volume} {298}},\ \bibinfo {pages} {502}
  (\bibinfo {year} {1985})}\BibitemShut {NoStop}%
\bibitem [{\citenamefont {{Samsing}}\ \emph {et~al.}(2017)\citenamefont
  {{Samsing}}, \citenamefont {{MacLeod}},\ and\ \citenamefont
  {{Ramirez-Ruiz}}}]{Samsingetal2017}%
  \BibitemOpen
  \bibfield  {author} {\bibinfo {author} {\bibfnamefont {J.}~\bibnamefont
  {{Samsing}}}, \bibinfo {author} {\bibfnamefont {M.}~\bibnamefont
  {{MacLeod}}},\ and\ \bibinfo {author} {\bibfnamefont {E.}~\bibnamefont
  {{Ramirez-Ruiz}}},\ }\bibfield  {title} {\bibinfo {title} {{Formation of
  Tidal Captures and Gravitational Wave Inspirals in Binary-single
  Interactions}},\ }\href {https://doi.org/10.3847/1538-4357/aa7e32} {\bibfield
   {journal} {\bibinfo  {journal} {\apj}\ }\textbf {\bibinfo {volume} {846}},\
  \bibinfo {eid} {36} (\bibinfo {year} {2017})},\ \Eprint
  {https://arxiv.org/abs/1609.09114} {arXiv:1609.09114 [astro-ph.HE]}
  \BibitemShut {NoStop}%
\bibitem [{\citenamefont {Naoz}(2016)}]{Naoz2016}%
  \BibitemOpen
  \bibfield  {author} {\bibinfo {author} {\bibfnamefont {S.}~\bibnamefont
  {Naoz}},\ }\bibfield  {title} {\bibinfo {title} {The eccentric kozai-lidov
  effect and its applications},\ }\href
  {https://doi.org/10.1146/annurev-astro-081915-023315} {\bibfield  {journal}
  {\bibinfo  {journal} {Annual Review of Astronomy and Astrophysics}\ }\textbf
  {\bibinfo {volume} {54}},\ \bibinfo {pages} {441} (\bibinfo {year} {2016})},\
  \Eprint
  {https://arxiv.org/abs/https://doi.org/10.1146/annurev-astro-081915-023315}
  {https://doi.org/10.1146/annurev-astro-081915-023315} \BibitemShut {NoStop}%
\bibitem [{\citenamefont {Hughes}(1995)}]{BarryHughes1995}%
  \BibitemOpen
  \bibfield  {author} {\bibinfo {author} {\bibfnamefont {B.~D.}\ \bibnamefont
  {Hughes}},\ }\href {https://doi.org/10.1079/PNS19950063} {\emph {\bibinfo
  {title} {Random walks and random environments. {V}ol. 1}}},\ Oxford Science
  Publications\ (\bibinfo  {publisher} {The Clarendon Press, Oxford University
  Press, Oxford},\ \bibinfo {year} {1995})\ pp.\ \bibinfo {pages} {xxii+631},\
  \bibinfo {note} {random walks}\BibitemShut {NoStop}%
\bibitem [{\citenamefont {{Heggie}}\ \emph {et~al.}(1996)\citenamefont
  {{Heggie}}, \citenamefont {{Hut}},\ and\ \citenamefont
  {{McMillan}}}]{Heggieetal1996}%
  \BibitemOpen
  \bibfield  {author} {\bibinfo {author} {\bibfnamefont {D.~C.}\ \bibnamefont
  {{Heggie}}}, \bibinfo {author} {\bibfnamefont {P.}~\bibnamefont {{Hut}}},\
  and\ \bibinfo {author} {\bibfnamefont {S.~L.~W.}\ \bibnamefont
  {{McMillan}}},\ }\bibfield  {title} {\bibinfo {title} {{Binary--Single-Star
  Scattering. VII. Hard Binary Exchange Cross Sections for Arbitrary Mass
  Ratios: Numerical Results and Semianalytic FITS}},\ }\href
  {https://doi.org/10.1086/177611} {\bibfield  {journal} {\bibinfo  {journal}
  {\apj}\ }\textbf {\bibinfo {volume} {467}},\ \bibinfo {pages} {359} (\bibinfo
  {year} {1996})}\BibitemShut {NoStop}%
\bibitem [{\citenamefont {{Press}}\ and\ \citenamefont
  {{Teukolsky}}(1977)}]{PressTeukolsky1977}%
  \BibitemOpen
  \bibfield  {author} {\bibinfo {author} {\bibfnamefont {W.~H.}\ \bibnamefont
  {{Press}}}\ and\ \bibinfo {author} {\bibfnamefont {S.~A.}\ \bibnamefont
  {{Teukolsky}}},\ }\bibfield  {title} {\bibinfo {title} {{On formation of
  close binaries by two-body tidal capture.}},\ }\href
  {https://doi.org/10.1086/155143} {\bibfield  {journal} {\bibinfo  {journal}
  {\apj}\ }\textbf {\bibinfo {volume} {213}},\ \bibinfo {pages} {183} (\bibinfo
  {year} {1977})}\BibitemShut {NoStop}%
\bibitem [{\citenamefont {{Kochanek}}(1992)}]{Kochanek1992}%
  \BibitemOpen
  \bibfield  {author} {\bibinfo {author} {\bibfnamefont {C.~S.}\ \bibnamefont
  {{Kochanek}}},\ }\bibfield  {title} {\bibinfo {title} {{The Dynamical
  Evolution of Tidal Capture Binaries}},\ }\href
  {https://doi.org/10.1086/170966} {\bibfield  {journal} {\bibinfo  {journal}
  {\apj}\ }\textbf {\bibinfo {volume} {385}},\ \bibinfo {pages} {604} (\bibinfo
  {year} {1992})}\BibitemShut {NoStop}%
\bibitem [{\citenamefont {{Fabian}}\ \emph {et~al.}(1975)\citenamefont
  {{Fabian}}, \citenamefont {{Pringle}},\ and\ \citenamefont
  {{Rees}}}]{Fabianetal1975}%
  \BibitemOpen
  \bibfield  {author} {\bibinfo {author} {\bibfnamefont {A.~C.}\ \bibnamefont
  {{Fabian}}}, \bibinfo {author} {\bibfnamefont {J.~E.}\ \bibnamefont
  {{Pringle}}},\ and\ \bibinfo {author} {\bibfnamefont {M.~J.}\ \bibnamefont
  {{Rees}}},\ }\bibfield  {title} {\bibinfo {title} {{Tidal capture formation
  of binary systems and X-ray sources in globular clusters.}},\ }\href
  {https://doi.org/10.1093/mnras/172.1.15P} {\bibfield  {journal} {\bibinfo
  {journal} {\mnras}\ }\textbf {\bibinfo {volume} {172}},\ \bibinfo {pages}
  {15} (\bibinfo {year} {1975})}\BibitemShut {NoStop}%
\bibitem [{\citenamefont {Hein\"{a}m\"{a}ki}\ \emph {et~al.}(1999)\citenamefont
  {Hein\"{a}m\"{a}ki}, \citenamefont {Lehto}, \citenamefont {Valtonen},\ and\
  \citenamefont {Chernin}}]{Heinamakietal1999}%
  \BibitemOpen
  \bibfield  {author} {\bibinfo {author} {\bibfnamefont {P.}~\bibnamefont
  {Hein\"{a}m\"{a}ki}}, \bibinfo {author} {\bibfnamefont {H.~J.}\ \bibnamefont
  {Lehto}}, \bibinfo {author} {\bibfnamefont {M.~J.}\ \bibnamefont
  {Valtonen}},\ and\ \bibinfo {author} {\bibfnamefont {A.~D.}\ \bibnamefont
  {Chernin}},\ }\bibfield  {title} {\bibinfo {title} {{Chaos in three-body
  dynamics: Kolmogorov—Sinai entropy}},\ }\href
  {https://doi.org/10.1046/j.1365-8711.1999.02859.x} {\bibfield  {journal}
  {\bibinfo  {journal} {Monthly Notices of the Royal Astronomical Society}\
  }\textbf {\bibinfo {volume} {310}},\ \bibinfo {pages} {811} (\bibinfo {year}
  {1999})},\ \Eprint
  {https://arxiv.org/abs/https://academic.oup.com/mnras/article-pdf/310/3/811/3135356/310-3-811.pdf}
  {https://academic.oup.com/mnras/article-pdf/310/3/811/3135356/310-3-811.pdf}
  \BibitemShut {NoStop}%
\bibitem [{\citenamefont {{Lichtenberg}}\ and\ \citenamefont
  {{Lieberman}}(1992)}]{LichtenbergLieberman1992}%
  \BibitemOpen
  \bibfield  {author} {\bibinfo {author} {\bibfnamefont {A.}~\bibnamefont
  {{Lichtenberg}}}\ and\ \bibinfo {author} {\bibfnamefont {M.}~\bibnamefont
  {{Lieberman}}},\ }\href@noop {} {\emph {\bibinfo {title} {{Regular and
  Chaotic Dynamics}}}}\ (\bibinfo  {publisher} {Springer-Verlag, New York},\
  \bibinfo {year} {1992})\BibitemShut {NoStop}%
\bibitem [{\citenamefont {Redner}(2001)}]{Redner2001}%
  \BibitemOpen
  \bibfield  {author} {\bibinfo {author} {\bibfnamefont {S.}~\bibnamefont
  {Redner}},\ }\href {https://doi.org/10.1017/CBO9780511606014} {\emph
  {\bibinfo {title} {A guide to first-passage processes}}}\ (\bibinfo
  {publisher} {Cambridge University Press, Cambridge},\ \bibinfo {year}
  {2001})\ pp.\ \bibinfo {pages} {x+312}\BibitemShut {NoStop}%
\end{thebibliography}%

\onecolumngrid
\appendix
\section{Marginal Energy Distribution -- Evaluation of The Angular Momentum Integral}
\label{appendix: angular momentum integral}
\begin{figure}
  \centering
  \includegraphics[width=0.9\textwidth]{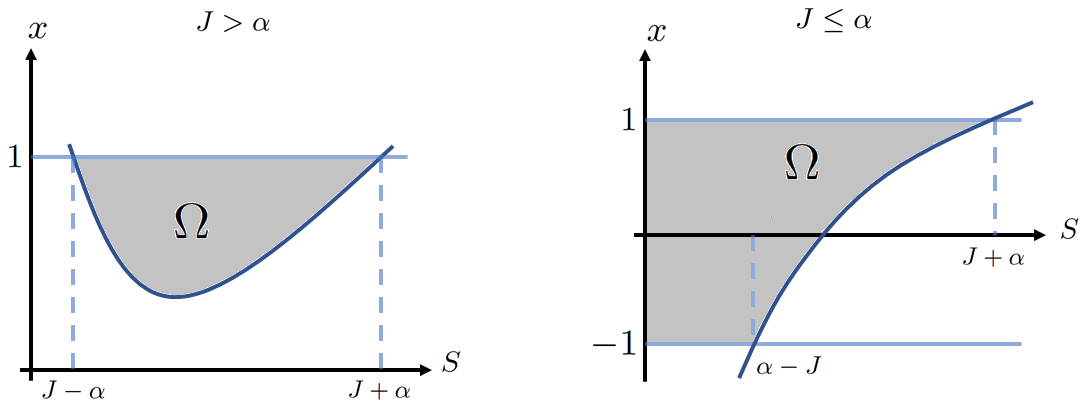}
  \caption{The two options for $\Omega$, with $t = \alpha$.}\label{fig:omega}
\end{figure}

Let us compute $\scri(t)$, as defined in equation \eqref{eqn: scri}, first approximating $\theta_{\max} = \theta_{ap}$.
There are two possibilities for $\Omega$: either $J > t$, or $J \leq t$; denote $\scri$ as $\scri_+$ and $\scri_-$, respectively, in these cases. Writing $J_b = S$, $J_a = Sx$, with $\abs{x}\leq 1$, one finds that in both cases $\Omega$ is a simple domain with respect to $x$, while $S$ is integrated from $0$ to $S_{\max} = \min\set{J+t,J_c}$. $\Omega$ is shown in figure \ref{fig:omega}.

For the case $t \leq J$, $\scri(t) = \scri_+(t)$, defined as
\begin{equation}
  \scri_+(t) = \frac{1}{J}\int_{J-t}^{S_{\max}}(t-\abs{J - S})\mathrm{d}S.
\end{equation}
Evaluating the last integral yields
\begin{equation}
  \scri_+(t) = \frac{1}{J}\begin{cases}
                 \frac{1}{2} \left(2 t (S_{\max}-J)-(J-S_{\max})^2+t^2\right), & \mbox{if } J<S_{\max} \land J - t \leq S_{\max}\\
                 \frac{1}{2} (-J+S_{\max}+t)^2, & \mbox{if } J\geq S_{\max} \land J - t \leq S_{\max} \\
                 0 & \mbox{otherwise}.
               \end{cases}
\end{equation}

For the case $t > J$: $\scri_- = \scri_-^{(1)} + \Theta(S_{\max} + J - t)\scri_-^{(2)}$, where $\Theta$ is the Heaviside theta function,
\begin{equation}
  \scri^{(1)}_-(t) = \frac{1}{J}\begin{cases}
                                  2J\min\set{t-J,S_{\max}}-J^2, & \mbox{if } \min\set{t-J,S_{\max}} \geq J \\
                                  \min\set{t-J,S_{\max}}^2, & \mbox{otherwise}.
                                \end{cases},
\end{equation}
and
\begin{equation}
  \scri^{(2)}_-(t) = \frac{1}{J}\int_{t-J}^{S_{\max}}(t-\abs{J - S})\mathrm{d}S.
\end{equation}
To compute $\scri^{(2)}_-$ one needs to consider the signs carefully. The result is
\begin{equation}
  \scri^{(2)}_-(t) = \frac{1}{J}\begin{cases}
                       - \frac{(t-J)^2}{2} + (t+J)S_{\max} - \frac{S_{\max}^2}{2} - J^2, & \mbox{if }   t-J \leq J<S_{\max} \\
                       (t+J)S_{\max} - \frac{S_{\max}^2}{2} - t^2 + J^2 + \frac{(t-J)^2}{2} , & \mbox{if } t > 2J  \\
                       (t-J)S_{\max} + \frac{S_{\max}^2}{2} - \frac{3(t-J)^2}{2}, & \mbox{if } J \geq S_{\max}.
                     \end{cases}
\end{equation}

Inserting $R$ from equation \eqref{eqn: R definition}, where $1 \lesssim \beta \lesssim 2$, the exact value of $\scri(\alpha(E_\bin))$ follows, roughly, a power-law until $J = \alpha$, where it begins to fall sharply (like $\scri_+$). The dominant contribution to $\scri_-$ is $\scri_-^{(1)}$, whence, when $\abs{E_\bin } \gg \abs{E}$, but $\alpha > J$, one may approximate $\scri$ by $\scri_-^{(1)}$, which goes like $\abs{E_\bin}^{-1}$ for high angular momentum, but like $\abs{E_\bin}^{-1/2}$ for low angular momentum.

In the unbound case, when $\abs{E_\bin } \gg \abs{E}$, one has $a_s \approx a_\bin m_s/\mu_\bin$, and
\begin{equation}
  \frac{A^2}{J_c^2} \approx \left(2\pm\frac{\beta\mu_\bin}{m_s}\right)\frac{\beta m_s\mu_s}{\mu_\bin^2}.
\end{equation}


An even better approximation would be to take into account the eccentricity dependence of $\theta_{\max}$, when computing $\scri$. The angle $\theta_{\max}$ is equal to $\theta_{ap}$ for $e_s = 1$, and it vanishes when $L^2$ saturates the periapsis bound of \S \ref{subsec: Celtica in angle-action variables}. Let us define $\xi=R/a_s$ and
\begin{equation}
  b^2 = \frac{L^2}{GM\mu_s^2R}\begin{cases}
          \frac{1}{2-R/a_s}, & \mbox{bound case} \\
          \frac{1}{2+R/a_s}, & \mbox{unbound case}.
        \end{cases},
\end{equation}
so that
\begin{equation}
  e_s^2 = \begin{cases}
          1-b^2\xi\left(2-\xi\right), & \mbox{bound case } \\
          1+b^2\xi\left(2+\xi\right), & \mbox{unbound case}.
        \end{cases}.
\end{equation}
Then $\theta_{\max}(b = 0) = \theta_{ap}$, while $\theta_{\max}(b = 1) = 0$.
One could write $\theta_{\max} = \theta_{ap}\frac{\theta_{\max}}{\theta_{ap}}$, and then approximate the fraction. While at $b = 0, \xi = 0$, this fraction is unity, it differs from 1 for non-zero $b$ even at $\xi =  0$. So, let us define expand $\frac{\theta_{\max}}{\theta_{ap}}$ in powers of $\xi$; keeping only the leading term we find
\begin{equation}
  \frac{\theta_{\max}}{\theta_{ap}} = \sqrt{1-b^2}(1+2b^2) + \sqrt{1-b^2}O(\xi),
\end{equation}
uniformly in $b$. We keep only the leading term, and, since we have shown above that $\scri \approx \scri_-^{(1)}$ for most values of $E_\bin$, except possibly those where $f_\bin$ is very small anyway, we will only compute a correction to $\scri_-^{(1)}$ here, but corrections to $\scri_+$ and $\scri_-^{(2)}$ may be obtained in a similar manner.
The integral we need to compute is
\begin{equation}
  \scri(t) = \iint_{\Omega} \frac{S\mathrm{d}S\mathrm{d}x}{L}\frac{\theta_{\max}}{\theta_{ap}} \approx \iint_{\Omega} \frac{S\mathrm{d}S\mathrm{d}x}{L}\sqrt{1-b^2}(1+2b^2).
\end{equation}
Changing variables from $x$ to $L$ gives
\begin{equation}
  \scri(t) = \frac{1}{J}\iint \mathrm{d}S\mathrm{d}L \sqrt{1-b^2}(1+2b^2).
\end{equation}
Fortunately, this integral may be computed analytically in terms of inverse trigonometric functions. Let
\begin{equation}
  \phi(u) = \frac{1}{20}\sqrt{1-u^2}(12+u^2(1+2u^2)) + \frac{3}{4}u\arcsin u,
\end{equation}
and let us focus on $\scri_-^{(1)}$. Performing the integrations and keeping in mind the limits (here, $S$ ranges from $0$ to $\min\set{t-J,S_{\max}}$ and $L$ goes from $J+S$ to $\abs{J-S}$), yields, for the bound case
\begin{equation}
  \frac{J}{A_p^2}\scri_{-,\textrm{bd}}^{(1)} = \phi\left(\frac{J+\min\set{t-J,S_{\max}}}{A_p}\right) - 2\phi(J/A_p) + \begin{cases}
                                                                                                  \phi\left(\frac{J-\min\set{t-J,S_{\max}}}{A_p}\right), & \mbox{if } \min\set{t-J,S_{\max}} \leq J \\
                                                                                                  2\phi(0) - \phi\left(\frac{\min\set{t-J,S_{\max}}-J}{A_p}\right), & \mbox{otherwise}.
                                                                                                \end{cases},
\end{equation}
where $A_p = \mu_s\sqrt{GMR}\sqrt{2-R/a_s}$. For the unbound case,
\begin{equation}
  \frac{J}{t^2}\scri_{-,\textrm{ubd}}^{(1)} = \phi\left(\frac{J+\min\set{t-J,S_{\max}}}{t}\right) - 2\phi(J/t) + \begin{cases}
                                                                                                  \phi\left(\frac{J-\min\set{t-J,S_{\max}}}{t}\right), & \mbox{if } \min\set{t-J,S_{\max}} \leq J \\
                                                                                                  2\phi(0) - \phi\left(\frac{\min\set{t-J,S_{\max}}-J}{t}\right), & \mbox{otherwise}.
                                                                                                \end{cases}.
\end{equation}

This improved, more cumbersome approximation differs significantly from the simpler one, made at the beginning of this section, when the masses are significantly different from each other. Therefore, we use it only for figures \ref{fig:Heggie masses} and \ref{fig:f_E errors} only, since only there are the masses considerably different or is there a need for a high degree of accuracy, and only there does it make a difference.

\section{Chaos As Phase-Space Diffusion}
\label{sec: chaos diffusion}

In this appendix we endeavour to give a heuristic justification of equation \eqref{eqn: sigma}, i.e. of the mixing assumption inside $\celtica$. We do so by defining $\celtica$ as the region in phase-space, in which the system is completely non-hierarchical. By energy considerations, the spatial extent of this region must be related to the binary's initial semi-major axis (recall that the binary is hard, so the amount of energy contributed by the in-coming star is negligible), so the spatial size of $\celtica$ must be approximately $R$, with $\beta$ of order unity. If the system is non-hierarchical inside $\celtica$, then, by dimensional analysis, the relevant time-scale must also be a function of the energy alone, i.e., it must be the virial time-scale $\tau_\textrm{vir} = G\sqrt{m_1m_2+m_1m_3+m_2m_3}M/(2\abs{E}^{3/2})$. Indeed, \citet{Heinamakietal1999} found that the Lyapunov time, $\lambda^{-1}$, is roughly
\begin{equation}
  \frac{1}{27\sqrt{6}}\frac{GM^{5/2}}{\abs{E}^{3/2}} = \frac{\sqrt{6}}{27}\tau_\textrm{vir}.
\end{equation}
for the equal mass case.\footnote{We assume that the initial longest distance between stars is $a_\bin$, and that the binary is hard.} While this indeed consolidates the assumption that $R$ is given by equation \eqref{eqn: R definition}, with $\beta$ of order unity, the weak dependence of $\sigma$ on $\beta$ in equation \eqref{eqn:binary cross-section} -- only through $\theta_{\max}$ -- implies that one cannot specify the precise value of $\beta$ from the simulations of ref. \cite{Heinamakietal1999} -- but this could be theoretically done with a very high resolution similar simulation. In this appendix we normalise the units of time by this time-scale, and the units of mass by the total mass (recall that all 3 bodies are taken to have masses of the same order of magnitude), such that both momenta and distances have units of length. We also assume for simplicity that all three masses are equal.

Outside $\celtica$, $f$ simply satisfies a Liouville equation
\begin{equation}\label{eqn:encounters Liouville}
  \frac{\partial f}{\partial t} = \pois{\ham,f},
\end{equation}
where now, one may write $\ham$ in the co-ordinates of the inner binary, and those of the two-body system formed by the outer body and the centre-of-mass of the inner binary. Both in $\aquitania$ and in $\belgica$, the Hamiltonian $\ham$ admits angle-action variables. Given an initial condition $f_0(\theta,J)$, the solution is
\begin{equation}
  f(\theta,J,t) = f_0(\theta - \Omega t,J).
\end{equation}

Let us assume that, in $\mathcal{C}$, the motion is practically stochastic. Motion is deterministic throughout the evolution, but the chaotic dependence on initial conditions implies that \emph{practically} it is random, on short times (cf. \citet{LichtenbergLieberman1992}). What we mean is, that if the system is inside a small region $\mathcal{R}$ in $\celtica$ of size $\eps$ at one instant, it may jump, at the following instant, to any place in $\exp(\lambda\delta t)\mathcal{R}$, where $\delta t$ is the time difference between the two instants, with $\lambda$ being the Lyapunov exponent.
This evolution is assumed to be valid for times that are of the same order as the Lyapunov time $\lambda^{-1}$, and as long as the system is in $\celtica$. Suppose that one starts with a phase-space distribution that is uniform on some $\eps$-neighbourhood of some $w_0 \in \celtica$, namely uniform in $\mathcal{R} = B_\eps(w_0) \subseteq \celtica$; then, after a time $t$, this phase-space density evolves to a uniform distribution in a sphere (in the metric in which the Lyapunov exponent is calculated) of volume $\sim V(B_\eps(w_0))e^{\lambda t}$.

Suppose now that we wish to start with an initial condition in $\celtica$ very close to a delta-function. Then, as time goes by, $f$ spreads over $\celtica$, until some parts of it reach $\celtica$'s boundaries, and enter $\aquitania$ or $\belgica$. The time-scale of evolution in $\celtica$ is roughly the virial time-scale of the three-body system, while the time-scales for interesting evolution in $\aquitania$ and $\belgica$ are set by the frequencies $\Omega$ of the outer binary. \emph{Per definitionem}, these are much longer than the virial time-scale, for otherwise the system would not be hierarchical. Thus, those parts of $f$ which have left $\celtica$, are, from the point of view of the $\celtica$, stuck at the boundary, so to speak. The distribution continues to spread, until virtually all of $f$ is on $\partial\celtica$. The proportion that arrives at $\aquitania$ never returns to $\celtica$, but the rest -- being in $\belgica$, eventually does return to $\celtica$, and starts all over. This goes on \emph{ad infinitum}, or until all of the probability mass is in $\aquitania$.

Separation of scales thus ensures that the evolution of the system proceeds as a sequence of close three-body, chaotic encounters, and between them -- hierarchical phases, until one of the three bodies is ejected. This picture meshes well with simulations \citep{Anosova1986,AnosovaOrlov1986,Samsingetal2014}, but here we have lent it some theoretical credence.

If $w_0$ is a typical initial condition inside $\celtica$, then its phase-space distance from the $\partial \celtica$ is roughly $R$. According to the evolution described above, the time it would take the system to arrive at $\partial \celtica$ is then
\begin{equation}
  t \sim \frac{1}{\lambda} \ln\left(\frac{R^8}{V(B_\eps(w_0))}\right) \sim \frac{\ln R - \ln \eps}{\lambda}.
\end{equation}
The power of $8$ is due to the conserved quantities: the three-body phase-space is 18-dimensional, but conservation of linear momentum in the centre-of-mass frame, angular momentum and energy reduce this dimension to $8$. Requiring a resolution $\eps/R \ll 1$ implies that the time it takes the system to leave $\celtica$ is larger than the Lyapunov time, consolidating the assumption of efficient phase-space mixing inside it. The simulations of ref. \cite{Manwadkaretal2020} demonstrated that the half-life time of chaotic three-body systems is
\begin{equation}
  2.6\times\frac{3}{\sqrt{2}} \approx 30.4 \lambda^{-1}
\end{equation}
(for the equal mass case, in units of $\tau_\textrm{vir}$, see their table 4). This is indeed sufficient for chaotic mixing.

The next step is to see how much of the probability mass arrives at each point in $\partial\celtica$ in a single close encounter. (In a close approach that is not the first one, by linearity, the solution is a superposition of solutions that start out as delta-functions around some point in $\celtica$.) Due to the scale-separation, the evolution equation for the distribution function in $\celtica$ should have (approximately) Dirichlet boundary conditions on $\partial\celtica$, whence the probability of reaching a point $w \in \partial \celtica$ -- which leads immediately to the outcome probability in $\aquitania$ and $\belgica$ via Liouville's equation -- is given by the so-called `eventual hitting probability' of $w$ \citep{Redner2001}, which is simply the time-integral of the scalar product of $\del f$ with the unit normal to $\partial \celtica$, $\hat{\mathbf{n}}$ (the gradient $\del$ is a phase-space gradient). As $f$ effectively evolves like a top hat that expands, $\del \int f\mathrm{d}t$ is very large at $\partial \celtica$, but its magnitude is independent of $w_0$. The scalar product $\hat{\mathbf{n}}\cdot\del f$ gives an additional cosine, so that
\begin{equation}
  f_\textrm{eventual}(w) \propto \cos\theta(w,w_0),
\end{equation}
where $\theta(w,w_0)$ is the angle between $\hat{\mathbf{n}}$ and the vector pointing from $w_0$ to $w$. However, the boundary between $\celtica$ and the other two sets is not well-defined, so instead, it would be useful to think of $\partial\celtica$ as a `fuzzy' boundary -- with some width. This implies that the cosine should be removed as un-physical, and replaced by its average $\int_{-\pi/2}^{\pi/2}\cos\theta \mathrm{d}\theta$. Thus, the eventual hitting probability is well-approximated by a function independent of both $w_0$ and $w$.

What one is interested in is the probability distribution of an outcome in $\aquitania$ or $\belgica$, and specifically, in the action distribution there. Consider the set $\mathcal{S}(J_a,J_b,J_c)$, which is the set of all points $w \in \aquitania \cup \belgica$ such that the (Delaunay) actions of the inner binary are between $(J_a,J_b,J_c)$ and $(J_a+\mathrm{d}J_a,J_b+\mathrm{d}J_b,J_c+\mathrm{d}J_c)$. What we are interested in is the measure of $\mathcal{S}(J_a,J_b,J_c)$. Given that these actions correspond to something the has been in a close triple system, one can use Liouville's theorem to find translate this question into a question of finding the probability distribution function of $\partial\celtica$. This is independent of the initial condition, which implies that one can simply compute it assuming a uniform distribution of initial conditions in $\celtica$, which we compute in \S \ref{sec: phase space integration}.

The reader should also bear in mind that due to the scale separation of the evolution in the hierarchical region $\belgica$, there is an additional source of phase-space mixing: while the outer body moves along its two-body orbit about the inner binary, the mean anomaly of the latter evolves much more rapidly, and its value when the third star returns depends on its initial value sensitively, which implies that it is effectively quite random. Therefore, as the number of close approaches increases, $\sigma$ should resemble equation \eqref{eqn: sigma} more closely -- this fact is attested to by the simulations of ref. \cite{StoneLeigh2019}. Indeed, if the above arguments apply only approximately, so that in a single close approach mixing is only efficient in a fraction $\nu$ of the total volume of $\celtica$, then the phase-space distribution still tends to a fully-mixed one, as $\nu^n$ -- exponentially in the number $n$ of close approaches. It may be possible that a close triple approach following a hierarchical phase end more quickly than is required for the system to mix chaotically in $\celtica$ \citep{Kol2020}. In that case, this `phase-mixing' implies that the cross-section is still mixed.

\section{Approximation Error Induced By Model}
\label{appendix:error estimates}
As the solution described in the previous sections is a statistical approximation to a deterministic, albeit chaotic, problem, it would be useful to be able to constrain the uncertainty due to the approximation above. We do so by evaluating $f_\bin$ for different values of $\beta$, and comparing the resulting plots. An analogous check may be performed for $\eta$.\footnote{The two are not independent, and in the plots we have kept $\eta\beta$ constant, when varying $\beta$; this affects only the low energy cut-off of the bound case.}
While this test is not an ideal uncertainty estimate, it is at least a test of the robustness of the model; we rely on the fact that in many situations, the error induced by varying the parameters of a model is of a similar size to the one made by using that model as an approximation.

\begin{figure}
  \centering
  \includegraphics[width=0.45\textwidth]{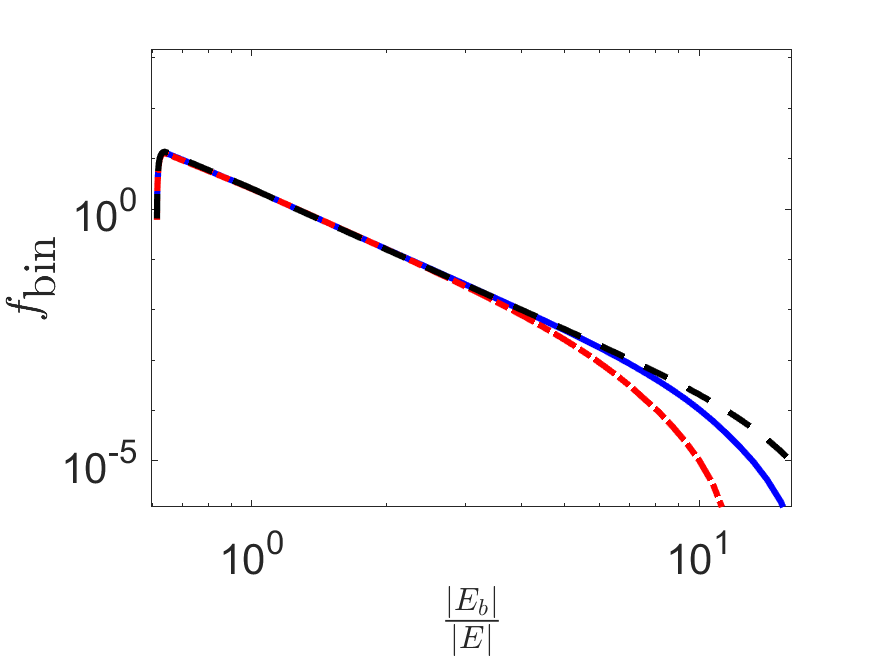}
  \caption{Plots of the marginal energy distribution (for $E_b = E_\bin$), in equation \eqref{eqn:f_E_bin marginal} for $m_1=m_2=m_3$ for different values of $\beta$: $\beta = 1.5$ (blue), $\beta = 1$ (red, dash-dotted), and $\beta = 2$ (black, dashed). One can see that except for at energies where $f_\bin$ is minuscule anyway, the three plots coincide.}\label{fig:f_E errors}
\end{figure}

\begin{figure*}
  \centering
  \includegraphics[width=0.325\textwidth]{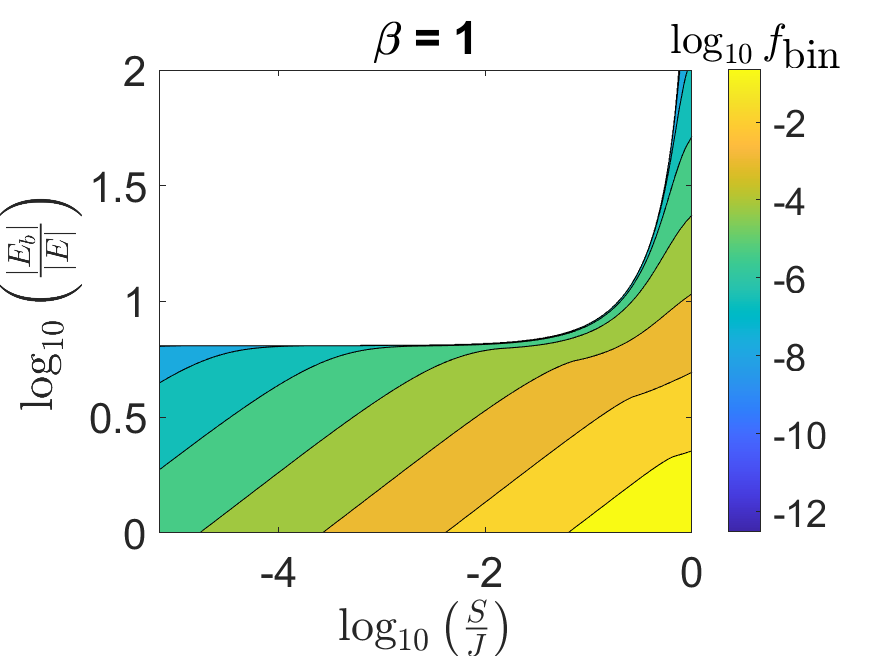}
  \includegraphics[width=0.325\textwidth]{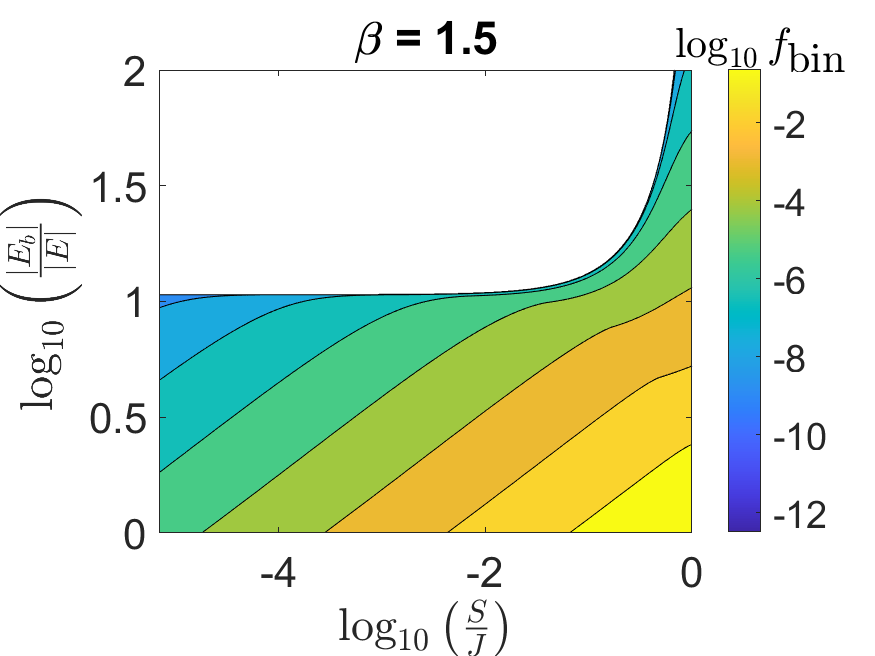}
  \includegraphics[width=0.325\textwidth]{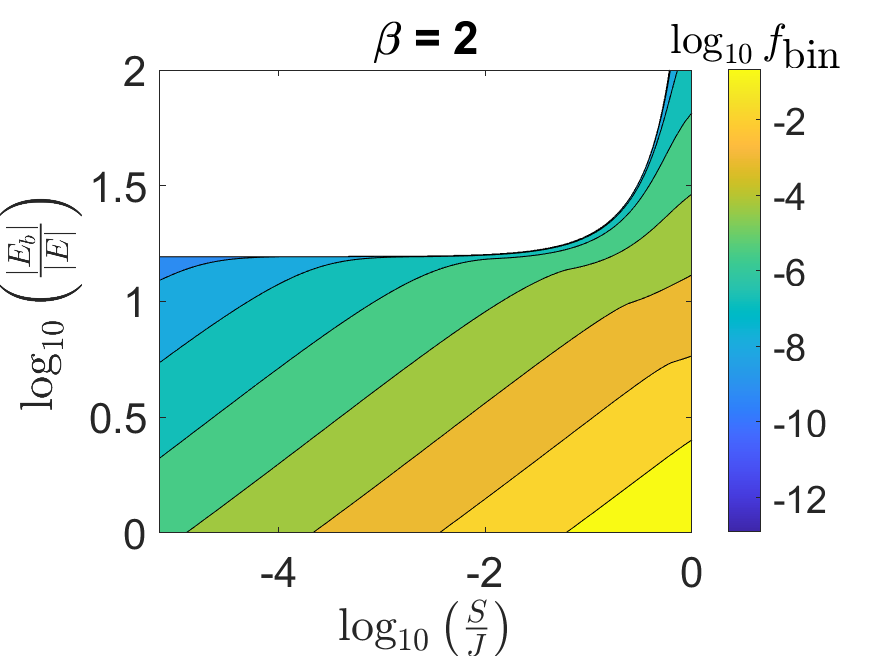}
  \caption{Plots of equation \eqref{eqn:f_bin masses} for $m_1=m_2=m_3$ for different values of $\beta$: $\beta = 1$ (left), $\beta = 1.5$ (centre), and $\beta = 2$ (right). One can see that except for at energies where $f_\bin$ is minuscule anyway, the three plots agree well with each other.}\label{fig:contour plots}
\end{figure*}

Figure \ref{fig:f_E errors} shows the marginal energy distribution, and figure \ref{fig:contour plots} shows contour plots of the joint distribution of $E_\bin$ and $S = \abs{\mathbf{S}}$; both figures indicate a weak dependence on $\beta$, and therefore a high accuracy of the approximation model made here.

\section{Inclination}
\label{appendix:inclination}
A yet more refined version of $R$ may be obtained by not approximating the Legendre polynomial by unity. Instead we use the singly-averaged correction to the Hamiltonian -- averaged over the inner binary's orbit. This yields a quadrupole term proportional to (see, e.g. ref. \cite{ValtonenKarttunen2006} for orbit-averaging procedures)
\begin{equation}
  \ham_{\textrm{quad}} = -\frac{GM_2}{8}\frac{a_b^2}{r_s^3}\left\{-3 \left(\cos^2i \left(e_b^2-1\right)+4 e_b^2+1\right) \cos (2\theta_b^s + \phi_s)+3 \cos^2i \left(e_b^2-1\right)-6 e_b^2+1\right\},
\end{equation}
where $\phi_s$ is the true anomaly of the outer orbit, $i$ is the mutual inclination between the two orbits,
\begin{equation}
  i = i_s + i_\bin = \arccos\left(\frac{J_z - S_z}{\abs{\mathbf{J}-\mathbf{S}}}\right) + \arccos\left(\frac{S_z}{S}\right),
\end{equation}
where $S = J_b$ and $S_z = J_a$ are the magnitude and $\zhat$-component of the binary angular momentum, respectively, and $\zhat$ is parallel to $\mathbf{J}$. We also shortened the sub-script $\bin$ to $b$ for brevity.

This quadrupole term is to be evaluated at the value $\theta_c^s$ that corresponds to $r_s = R$, i.e. at the value of $\phi_s$ that corresponds to it, then equated with $E$, and solved for $R$. This procedure would give an $R$ that depends on the binary actions and on $\theta_b^s$; but, as we already know that the dependence of $\theta_{\max}$ on the precise value of $R$ is weak, we neglect this dependence on $\theta_b^s$, by setting $\ham_{\textrm{quad}}$ to its average value, when averaging it over $\theta_b^s$, thereby allowing us to solve for $R$, explicitly:
\begin{equation}\label{eqn: R inclination}
  R = \beta\min\set{a_\bin, \frac{a_\bin^{2/3}}{2}\frac{GM_2}{\abs{E}}\left|3 \left(e_\bin^2-1\right) \cos ^2(i)-6 e_\bin^2+1\right|^{1/3}}.
\end{equation}

\begin{figure*}
  \centering
  \includegraphics[width=0.45\textwidth]{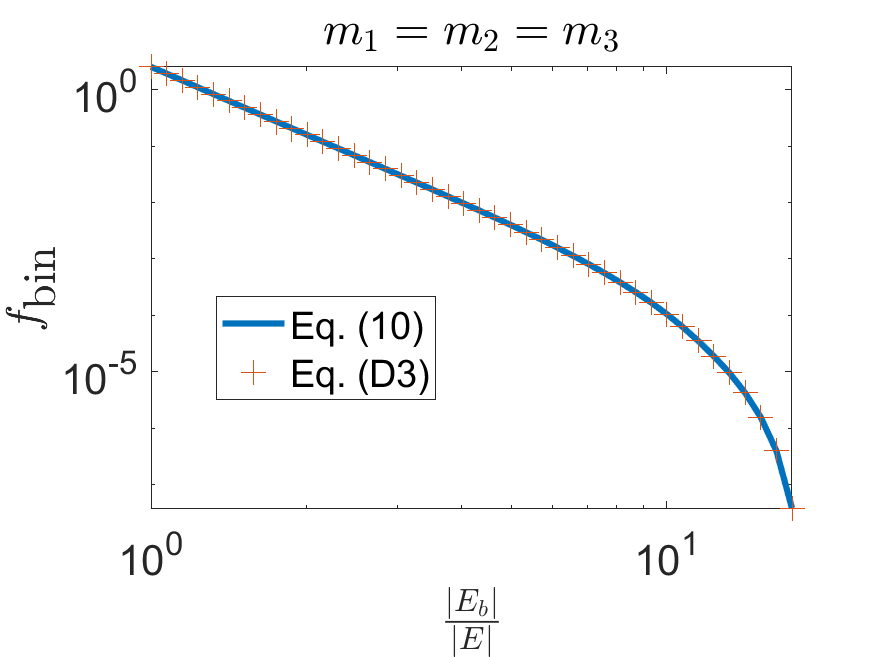}
  \includegraphics[width=0.45\textwidth]{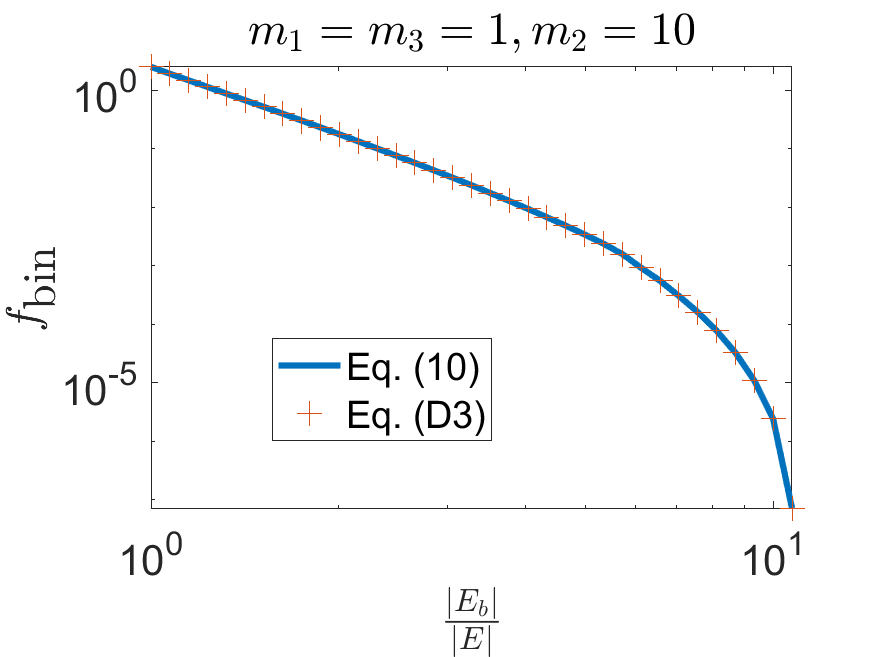}
  \includegraphics[width=0.45\textwidth]{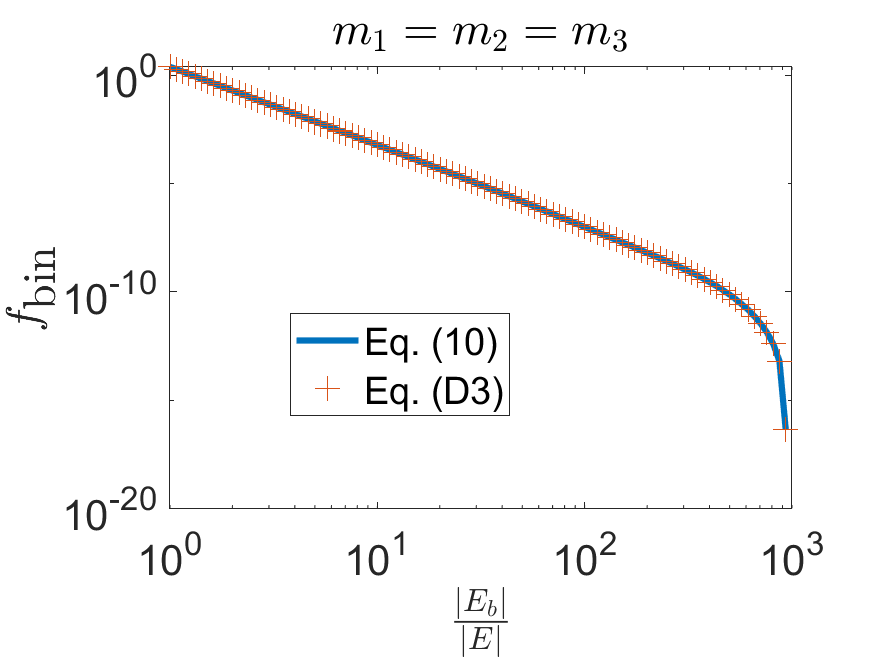}
  \includegraphics[width=0.45\textwidth]{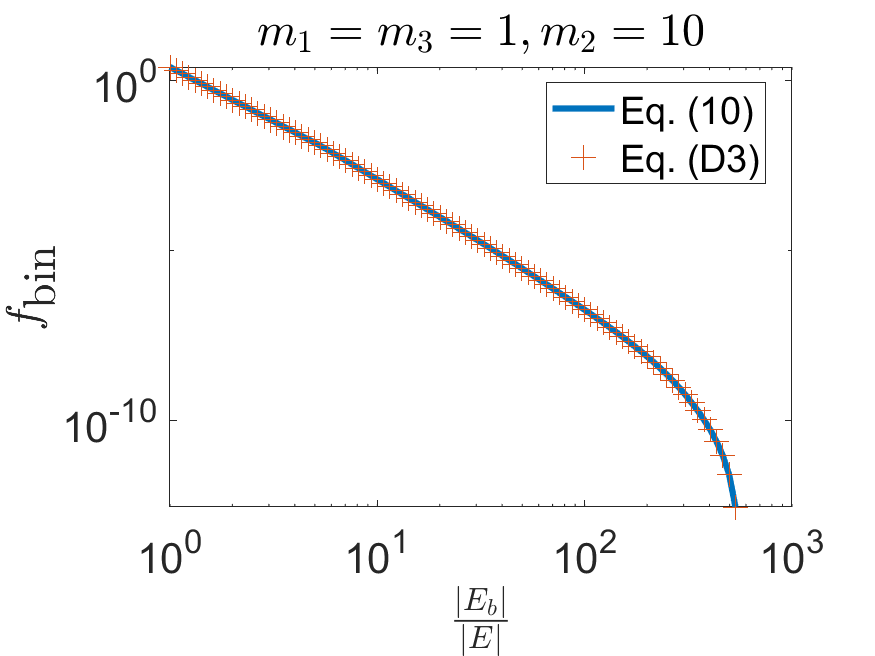}
  \includegraphics[width=0.45\textwidth]{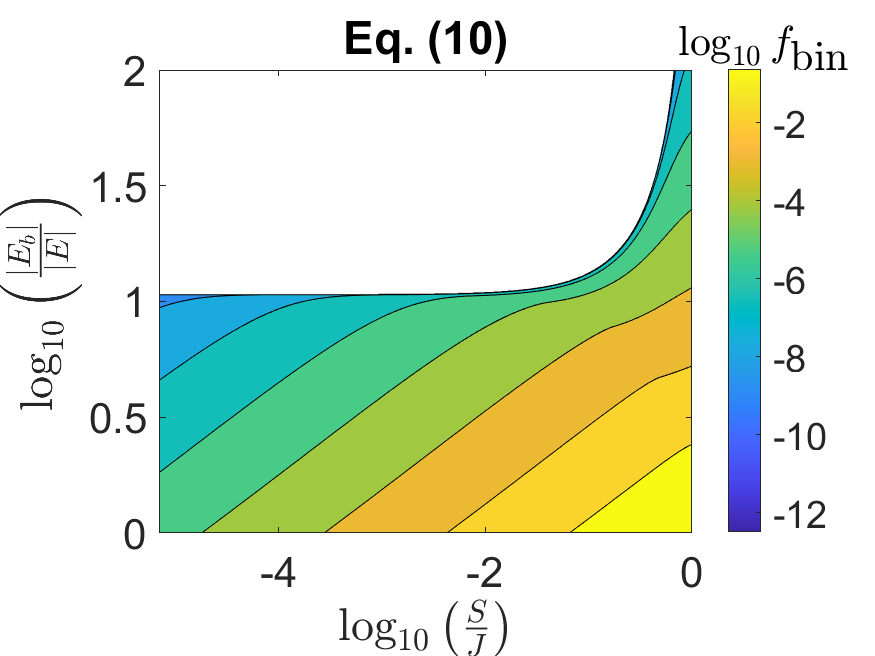}
  \includegraphics[width=0.45\textwidth]{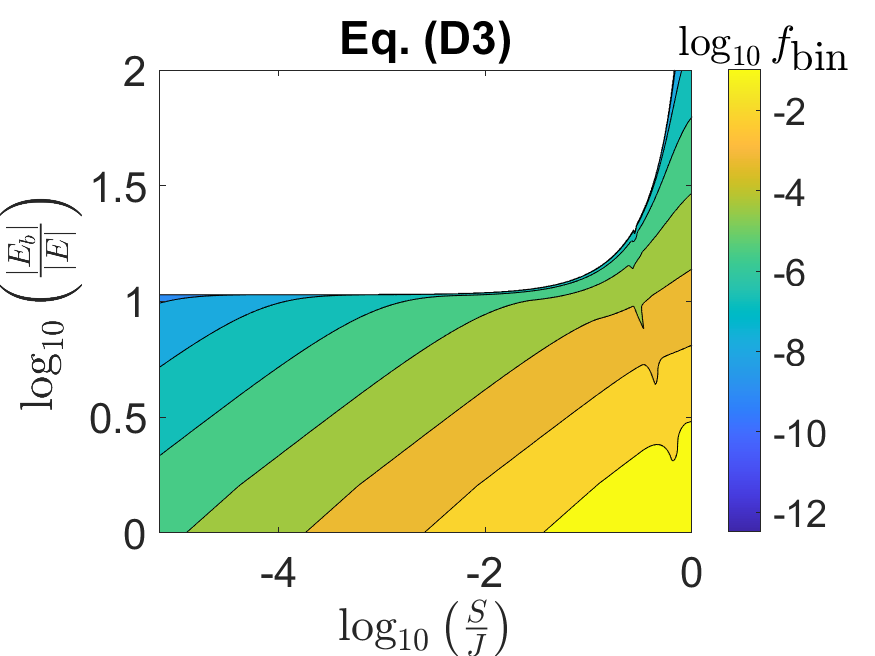}
  \caption{A comparison between $f_\bin$, calculated with $R$ defined by equation \eqref{eqn: R definition}, and by equation \eqref{eqn: R inclination}. The first two rows show the marginal energy distribution, and the last line shows that joint energy-spin distribution (left: equation \eqref{eqn: R definition}; right: equation \eqref{eqn: R inclination}). The top row displays $f_\bin$ for a large value of $J$, while the second -- for a low value. For the bottom row we used the same value of $J$ as for the top row, and equal masses. All rows have $\beta = 1.5$.}\label{fig:inclination R}
\end{figure*}

In figure \ref{fig:inclination R} we display a comparison between $f_\bin$ for $R$ given by equation \eqref{eqn: R definition}, and the correction implies by using instead equation \eqref{eqn: R inclination} (for the unbound case). The two definitions yield almost identical values of $f_\bin$ and therefore we use the simpler equation \eqref{eqn: R definition} in the paper.

\end{document}